\documentclass[a4paper,11pt]{article}
\usepackage{jheppub}
\usepackage{amsmath,amssymb,amsfonts}
\usepackage{a4wide,epsfig,psfrag,scalefnt}
\usepackage[dvipsnames]{xcolor}
\usepackage{braket}
\usepackage{placeins}
\usepackage{breqn}
\usepackage{slashed}
\usepackage{enumitem}
\usepackage[numbers,sort&compress]{natbib}
\usepackage{caption}
\usepackage{subcaption}
\usepackage{verbatim}
\usepackage{orcidlink}
\usepackage{amsmath}
\usepackage{float}
\usepackage{placeins}
\usepackage{afterpage}

\graphicspath{ {figs/} }
\newcommand{\ep}{\epsilon}

\def\gev{{\rm GeV}}

\newcommand{\spp}{\vphantom{$\Big($}}

\newcommand{\eqn}[1]{eq.\,(\ref{#1})}
\newcommand{\eqs}[1]{eqs.\,(\ref{#1})}

\newcommand{\fig}[1]{fig.\,\ref{#1}}

\newcommand{\citere}[1]{ref.~\cite{#1}}
\newcommand{\citeres}[1]{refs.~\cite{#1}}

\newcommand{\ord}[1]{{\cal O}(#1)}

\newcommand{\eqa}[1]{\begin{eqnarray} #1 \end{eqnarray}}

\newcommand{\C}[1]{{\cal{C}}_{#1}}
\newcommand{\MSbar}{$\overline{\text{MS}}$}
\newcommand{\qfto}{\mathcal{O}}
\newcommand{\eps}{{\color{black}\epsilon}}
\newcommand{\epssq}{{\color{black}\epsilon^2}}
\newcommand{\epstrip}{{\color{black}\epsilon^3}}

\allowdisplaybreaks

\title{
\boldmath Multi-parton contributions to $\bar B \to X_s \gamma$ at NLO}

\author{Kevin Brune \orcidlink{0000-0003-1048-3918},}
\author{Tobias Huber \orcidlink{0000-0002-3851-0116},}
\author{Lars-Thorben Moos \orcidlink{0000-0001-9398-4456}}
\affiliation{
Theoretische Physik 1, Center for Particle Physics Siegen (CPPS), Universit\"at Siegen, \\ Walter-Flex-Stra{\ss}e 3, D-57068 Siegen, Germany
}

\emailAdd{brune@physik.uni-siegen.de}
\emailAdd{huber@physik.uni-siegen.de}
\emailAdd{ltmoos@t-online.de}

\abstract{
Many contributions to the decay rate of the inclusive radiative $\bar{B}\rightarrow X_s \gamma$ transition have been calculated to NNLO in QCD during the past decades. However, there are still a few unknown contributions from multi-parton final states which are formally NLO. In the present work, we compute those four-body $b \rightarrow s\, q\, \bar{q}\,\gamma$ contributions at NLO in QCD which need to be supplemented by the five-body $b \rightarrow s\, q\, \bar{q}\, g\,\gamma $ bremsstrahlung. This calculation formally completes the purely perturbative contributions to $\bar{B}\rightarrow X_s \gamma$ at NLO. Our results are obtained by applying modern techniques of integral reduction and evaluation of master integrals. In particular, the analytic integration over the four and five-particle phase space in the presence of a cut on the photon energy turns out to be technically involved. We give our results completely analytically in terms of multiple polylogarithms, including the dependence on the collinear logarithms which arise from the mass-regularisation of collinear divergences. The numerical impact of multi-parton corrections on the $\bar{B}\rightarrow X_s \gamma$ decay rate turns out to be small, owing to a partial cancellation between LO and NLO contributions.
}

\preprint{
\begin{minipage}{3cm}
\small
\flushright
SI-HEP-2025-22\\
P3H-25-068 \\

\end{minipage}}

\begin{document}

\maketitle

%%%%%%%%%%%%%%%%%%%%%%%%%%%%%%%%%%%%%%%%%%%%%%%%%%%%%%%%%%%%%%%%%%%%%
\section{Introduction}
\label{Sec:Intro}
%%%%%%%%%%%%%%%%%%%%%%%%%%%%%%%%%%%%%%%%%%%%%%%%%%%%%%%%%%%%%%%%%%%%%

In the absence of a clear signal of physics beyond the Standard Model (SM), the current paradigm in high-energy physics is to achieve precise experimental measurements that are matched to very accurate theoretical predictions, with the goal of disentangling possible signs of new phenomena from uncertainties arising from SM effects. In this respect the inclusive radiative decay $\bar{B} \to X_s \gamma$ constitutes one of the most precise tests of the SM in the quark flavour sector and represents a standard candle in the indirect search for New Physics. At the parton level, the primary contribution comes from the two-body $b \rightarrow s \gamma$ decay, which is forbidden at tree-level in the SM due to its flavour-changing neutral current nature, and therefore highly sensitive to virtual contributions from new particles running in a loop. In this way, one can use low-energy observables to probe scales that are as high or --~depending on the process~-- even higher than those accessible via on-shell production of new degrees of freedom.

On the experimental side the value for the CP- and isospin-averaged branching ratio ${\mathcal{B}}_{s\gamma}$ of $\bar{B} \to X_s \gamma$ has been measured to $\pm 5.4\%$ accuracy. The current experimental average is given by~\cite{HFLAV:2022esi}
\begin{equation}
    \mathcal{B}_{s\gamma}^{\text{exp}} \, = \, (3.49\pm0.19)\times 10^{-4},
\end{equation}
once extrapolated to a photon-energy cut of $E_{\gamma} > E_0 = 1.6 \text{ GeV}$. This number has to be supplemented by a theoretical prediction where the uncertainty ideally is of a comparable size. In the past two decades a lot of work has been done to achieve a similar precision for the theoretical prediction, for the major updates see, e.g.,~\citeres{Gambino:2001ew,Misiak:2006zs,Misiak:2015xwa,Misiak:2020vlo}. The current SM prediction for $E_0 = 1.6 \text{ GeV}$ is given by~\cite{Misiak:2020vlo} 
\begin{equation}
     \mathcal{B}_{s\gamma}^{\text{SM}}=(3.40\pm0.17)\times 10^{-4},
\end{equation}
which includes corrections up to next-to-next-to-leading order (NNLO) in the strong coupling $\alpha_s$. The total uncertainty of $\pm 5\%$ is comprised of $\pm 3\%$ from higher-order effects, $\pm 3\%$ from interpolation in $m_c$, and $\pm 2.5\%$ from parametric and non-perturbative uncertainties, added in quadrature~\cite{Misiak:2020vlo}. With the upcoming runs of Belle II and the combination with data from the $B$-factories, the uncertainty on the experimental side is envisaged to decrease to the $\pm 2.6\%$ level~\cite{Belle-II:2018jsg,Ishikawa:2019TalkLyon}, calling for an increased effort also on the theory side. For recent progress on the perturbative and power-correction side, see for instance~\cite{Greub:2023msv,Fael:2023gau,Czaja:2023ren,Greub:2024mwp} and~\cite{Gunawardana:2019gep,Benzke:2020htm,Hurth:2023eqt,Bartocci:2024bbf}, respectively.

Despite the fact that many NNLO corrections are already available, even the next-to-leading order (NLO) corrections at leading power in the operator product expansion of $\bar B \to X_s \gamma$ are not completely known. At the partonic level, there are multi-parton final states of the form $b \to s \, q \, \bar q \, \gamma \, (g)$, whose leading-order (LO) tree-level contributions were calculated in~\cite{Kaminski:2012eb}. In~\citere{Huber:2014nna} those NLO contributions that require four-body $b \to s \, q \, \bar q \, \gamma$ final states only were obtained. In the present work, we compute those one-loop four-particle $b \to s \, q \, \bar{q} \, \gamma$ diagrams that must be supplemented by the corresponding five-particle tree-level cuts $b \to s \, q \, \bar{q} \, g \, \gamma $ from gluon bremsstrahlung. They are suppressed by small CKM factors or Wilson coefficients, but must be included to make the NLO part of the purely perturbative calculation formally complete. In this regard, the paper at hand is the last piece that is missing in order to formally complete $\bar{B} \rightarrow X_s \gamma$ at NLO in QCD at leading power. 

The calculation of the $b \to s \, q \, \bar q \, \gamma \, (g)$ contribution turns out to be technically involved. It amounts to computing interferences between current-current and QCD penguin operators from the effective weak Hamiltonian. Here, we will put special emphasis on the treatment of $\gamma_5$ in dimensional regularisation. The one-loop four-body and tree-level five-body integrals are subsequently processed using modern techniques of integral reduction and reverse unitarity. The master integrals are then computed using differential equations and explicit analytic integration over the $D$-dimensional four- and five-particle phase space (for previous work on multi-particle phase-space integrations see, e.g.,~\cite{Achasov:2003re,Heinrich:2006sw,Fael:2016yle,Pruna:2016spf,Cata:2016epa,Gituliar:2018bcr,Huber:2018gii,Magerya:2019cvz,Magerya:2025qgp}). The subsequent ultraviolet (UV) renormalisation and infrared (IR) subtraction renders our expressions finite in the dimensional regulator. However, the translation of dimensionally regulated collinear divergences into mass-regulated ones introduces collinear logarithms of the light quark masses. Finally, the numerical impact of our analytic results is investigated.

The content of this paper is as follows. In section~\ref{Sec:Theory} we discuss the theoretical framework of the calculation and introduce the different building blocks which have to be calculated. This is followed by discussing the bare calculation in section~\ref{Sec:barecalculation}. Here we discuss the diagrams and operator insertions, the treatment of $\gamma_5$, the integral reduction, phase-space integration, and outline the methods for computing the master integrals. In section~\ref{sec:renormalisation} we discuss UV renormalisation and the procedure to subtract the remaining IR divergences. Finally, in sections~\ref{sec:analyticresults} and~\ref{sec:numericalresults} we give analytic results for the different operator insertions and investigate the numerical size of the computed correction, respectively. We conclude in section~\ref{sec:conclusion} and relegate relations between phase-space integrals to appendix~\ref{app:integralrelations}.

%%%%%%%%%%%%%%%%%%%%%%%%%%%%%%%%%%%%%%%%%%%%%%%%%%%%%%%%%%%%%%%%%%%%%
\section{Theoretical Framework}
\label{Sec:Theory}
%%%%%%%%%%%%%%%%%%%%%%%%%%%%%%%%%%%%%%%%%%%%%%%%%%%%%%%%%%%%%%%%%%%%%
We work in the framework of the effective weak theory that is obtained from the SM by the decoupling of the $W$-bosons and all heavier particles. The relevant interaction terms for this work are given by the following Lagrangian
\begin{equation}
     \label{eq:Leff}
     \mathcal{L}_{\rm eff}=\mathcal{L}_{\rm QED+QCD}+\frac{4 G_F}{\sqrt{2}} V_{ts}^* V_{tb}
\left[  \sum_{i=1}^2 (\C{i}^u P_i^u + \C{i}^c P_i^c) + \sum_{i=3}^8 \C{i} P_i  \right]        +\rm{h.c.}\, ,
\end{equation}
where $\mathcal{L}_{\rm QED+QCD}$ is the standard QED and QCD part from the SM, and $\mathcal{C}_i^u$ and $\mathcal{C}_i$ denote the Wilson coefficients associated with the corresponding effective operators defined in~\citere{Chetyrkin:1996vx}
\begin{alignat}{4}\label{eq:operators}
\allowdisplaybreaks
P_1^u &= (\bar{s}_L \gamma_{\mu} T^a u_L)(\bar{u}_L \gamma^{\mu} T^a b_L) \,,\nonumber &&  
P_2^u &&= (\bar{s}_L \gamma_{\mu} u_L)(\bar{u}_L \gamma^{\mu} b_L) \,, \\[0.3em]
P_1^c &= (\bar{s}_L \gamma_{\mu} T^a c_L)(\bar{c}_L \gamma^{\mu} T^a b_L) \,,\nonumber &&  
P_2^c &&= (\bar{s}_L \gamma_{\mu} c_L)(\bar{c}_L \gamma^{\mu} b_L) \,, \\[0.3em]
P_3 &= (\bar{s}_L \gamma_{\mu} b_L)\textstyle{\sum_{q}}(\bar{q} \gamma^{\mu} q) \,, && 
P_4 &&= (\bar{s}_L \gamma_{\mu} T^a b_L)\textstyle{\sum_{q}}(\bar{q} \gamma^{\mu} T^a q)\,,  \nonumber \\[0.3em]
P_5 &= (\bar{s}_L \gamma_{\mu}\gamma_{\nu}\gamma_{\rho} b_L)\textstyle{\sum_{q}}(\bar{q} \gamma^{\mu}\gamma^{\nu}\gamma^{\rho} q)\,, \quad && 
P_6 &&= (\bar{s}_L \gamma_{\mu}\gamma_{\nu}\gamma_{\rho} T^a b_L)\textstyle{\sum_{q}}(\bar{q} \gamma^{\mu}\gamma^{\nu}\gamma^{\rho} T^a q) \,, \nonumber \\[0.3em]
P_7 &= \displaystyle \frac{e}{16\pi^2} m_b (\bar s_L\sigma^{\mu\nu} b_R)F_{\mu\nu}\, ,
&& P_8&&= \displaystyle \frac{g_s}{16\pi^2} m_b (\bar s_L\sigma^{\mu\nu} T^a b_R)G^a_{\mu\nu}\, .
\end{alignat}
Note that we follow the convention of~\citere{Huber:2014nna} such that $\C{1,2}^{q}$ ($q=u,c$) contain CKM phases, $\C{1,2}^{q} = -\lambda_q C_{1,2}$, with $\lambda_q\equiv V_{qs}^* V_{qb}/V_{ts}^* V_{tb}$, and the other Wilson coefficients are simply $\mathcal{C}_{3,\dots,6} = C_{3,\dots,6} $.

\begin{figure}[t]
\begin{minipage}[b]{.33\linewidth}
\centering
\vspace{0.35cm}
\subfloat[{}]{\vspace*{0.05cm}\includegraphics[scale=.42]{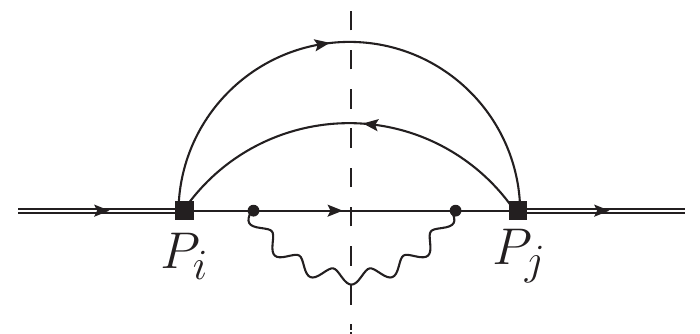}}
\end{minipage}%
\begin{minipage}[b]{.33\linewidth}
\centering
\subfloat[{}]{\includegraphics[scale=.42]{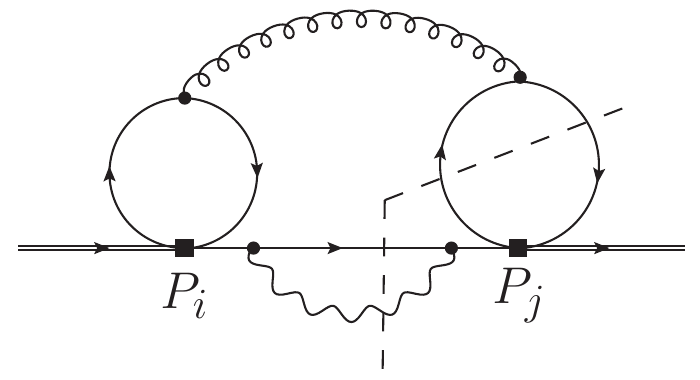}}
\end{minipage}
\centering
\begin{minipage}[b]{.33\linewidth}
\centering
\subfloat[{}]{\vspace*{0.1cm}\includegraphics[scale=.42]{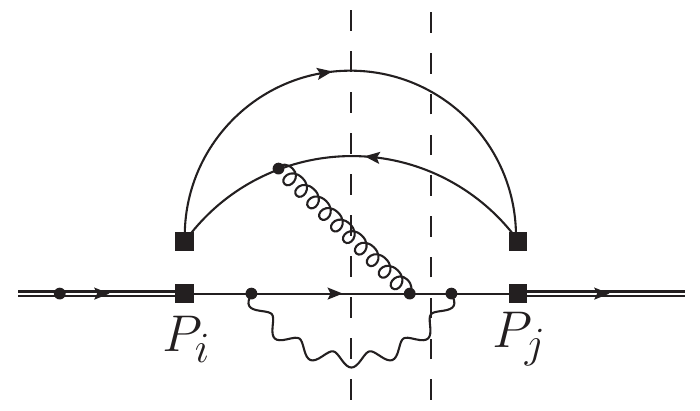}}
\end{minipage}
\caption{Sample cut-diagrams contributing to LO and NLO four- and five-body contributions. In this work we will focus on contributions from panel (c).}\label{fig:SampleDiaHatGij}
\end{figure}

As stated in the introduction the goal of this work is to formally complete the multi-parton contributions to $\bar{B} \rightarrow X_s \gamma$ at NLO in QCD. The inclusive decay rate can be written as
\begin{equation}
    \label{eq:DecayRate}
    \Gamma(\bar{B} \rightarrow X_s \gamma)_{E_{\gamma}>E_0}=\Gamma(b \rightarrow X_s^{\rm parton} \gamma)_{E_{\gamma}>E_0}+\mathcal{O}(\Lambda_{\rm{QCD}}/m_b)\,,
\end{equation}
where the first term on the right-hand side (RHS) is the inclusive partonic decay rate of the $b$ quark at leading power and the second term comprises power corrections. $ X_s^{\rm parton}$ denotes a state of quarks, anti-quarks and gluons with a net baryon number of $1/3$, a total strangeness of $S=-1$ and excluding charm. $E_{\gamma}>E_0$ means that the photon energy must be above a cut energy $E_0$. We are further allowed to expand the partonic decay rate into the different final states\footnote{We tacitly assume the cut on the photon energy to be present on the RHS as well.}, i.e.
\begin{equation}
    \label{eq:PartonicDecayRateExpand}
    \Gamma(b \rightarrow X_s^{\rm parton} \gamma)_{E_{\gamma}>E_0}=\Gamma(b \rightarrow s \gamma)+\ldots+\Gamma(b \rightarrow s q\bar{q}\gamma)+\Gamma(b \rightarrow s q\bar{q}g\gamma)+\ldots \, ,
\end{equation}
where in the present work we focus on the indicated four- and five-body contributions. The full partonic decay rate can be written, following the notation of~\citere{Huber:2014nna}, as
\begin{equation}
    \label{eq:PartonicDecayRateFull}
    \Gamma(b \rightarrow X_s^{\rm parton} \gamma)_{E_{\gamma}>E_0}= \Gamma_0 \sum_{i,j} \mathcal{C}_i^{\text{eff} \, *}(\mu) \, \mathcal{C}_j^{\text{eff}}(\mu) \,  \widetilde{G}_{ij}(\mu,z_c,\delta) \,,
\end{equation}
where we introduced the ``effective'' Wilson coefficients $\C{1q,2q,3,\ldots,6}^{\text{eff}} = \C{1q,2q,3,\ldots,6}$, $\C7^{\text{eff}} = \C7- \frac13 \C3 - \frac49 \C4 - \frac{20}3 \C5 - \frac{80}9 \C6$ and $\C8^{\text{eff}} = \C8+\C3 - \frac16 \C4 + 20 \C5 - \frac{10}3 \C6$ and the normalisation factor
\begin{equation}
\label{eq:normalisationfactor}
\Gamma_0 = \frac{G_F^2 m_b^5 \alpha_e |V_{ts}^{*}V_{tb}|^2}{32 \pi^4} \,.
\end{equation}
The arguments $z_c=m_c^2/m_b^2$ and $\delta$ denote dependence on the charm-quark mass, which appears in internal loops~\cite{Huber:2014nna}, and the photon energy cut via $\delta = 1-2E_0/m_b$, respectively. The four- and five-body contributions in question are given by
\begin{equation}
    \label{eq:PartonicDecayRate4B5B}
    \Gamma(b \rightarrow s q\bar{q}\gamma)_{E_{\gamma}>E_0}+\Gamma(b \rightarrow s q\bar{q}g\gamma)_{E_{\gamma}>E_0}= \Gamma_0 \sum_{i,j} \mathcal{C}_i^{\text{eff} \, *}(\mu) \, \mathcal{C}_j^{\text{eff}}(\mu) \,  \widehat{G}_{ij}(\mu,z_c,\delta) \,.
\end{equation}
The objects $\widehat{G}_{ij}(\mu,z_c,\delta)$ arise from the interference of operators $P_i$ and $P_j$ given in eqs.~\eqref{eq:operators} integrated over phase space. Furthermore, the matrix $\widehat{G}_{ij}(\mu,z_c,\delta)$ has a perturbative expansion in the strong coupling $\alpha_s$,
\begin{equation}
\label{eq:defGij}
 \widehat{G}_{ij}(\mu,z_c,\delta)= \widehat{G}_{ij}^{(0)}(\delta)+ \frac{\alpha_s(\mu)}{4 \pi} \widehat{G}_{ij}^{(1)}(\mu,z_c,\delta) + \mathcal{O}(\alpha_s^2) \,.
\end{equation}
The leading contribution $\widehat{G}_{ij}^{(0)}(\delta)$ to this matrix  (fig.~\ref{fig:SampleDiaHatGij}(a)) has been calculated in~\citere{Kaminski:2012eb}. The sub-leading contribution $\widehat{G}_{ij}^{(1)}(\mu,z_c,\delta)$ is split up according to $\displaystyle\widehat{G}_{ij}^{(1)}(\mu,\delta) \equiv {G}_{ij}^{(1)}(\mu,z_c,\delta)+{G}_{ij}^{(1)}(\mu,\delta)$. By definition, the first term includes all NLO contributions computed in~\cite{Huber:2014nna}, i.e.\ those results where the four-body contributions renormalise among themselves, without the inclusion of five-body contributions (fig.~\ref{fig:SampleDiaHatGij}(b)). The second term ${G}_{ij}^{(1)}(\mu,\delta)$ comprises the remaining diagrams sketched in figure~\ref{fig:SampleDiaHatGij}(c) where the four-body virtual contributions have to be supplemented by five-body real-radiation diagrams. ${G}_{ij}^{(1)}(\mu,\delta)$ has so far been unknown and is the main goal of the present work.

\begin{figure}[t!]
\centering
\begin{minipage}{.45\linewidth}
\centering
\subfloat[{}]{\includegraphics[scale=.55]{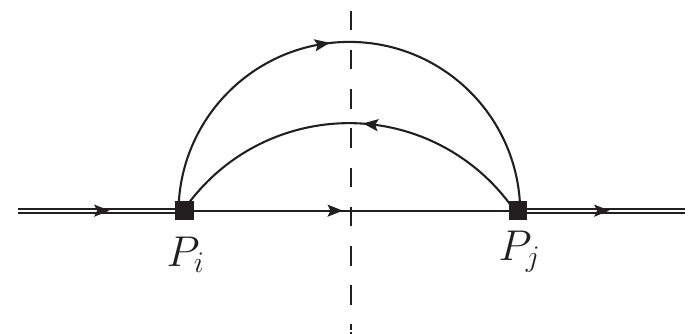}}
\end{minipage}
\begin{minipage}{.45\linewidth}

\subfloat[{}]{\includegraphics[scale=.55]{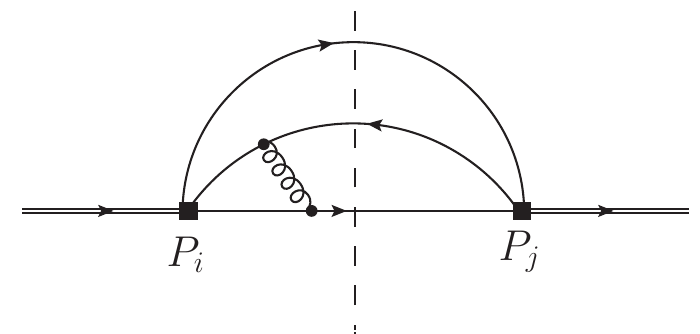}}
\end{minipage}
\\[1.0em]
\begin{minipage}{.45\linewidth}
\centering
\subfloat[{}]{\includegraphics[scale=.55]{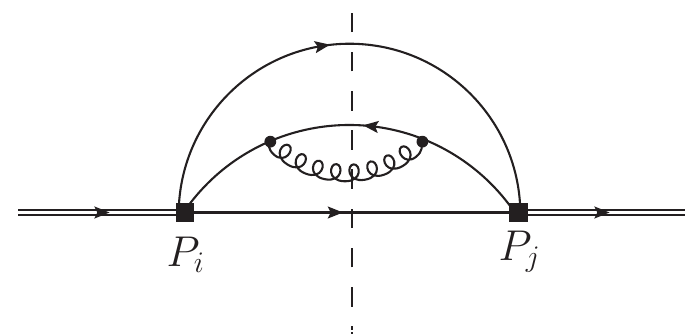}}
\end{minipage}
\begin{minipage}{.45\linewidth}
\centering
\subfloat[{}]{\includegraphics[scale=.55]{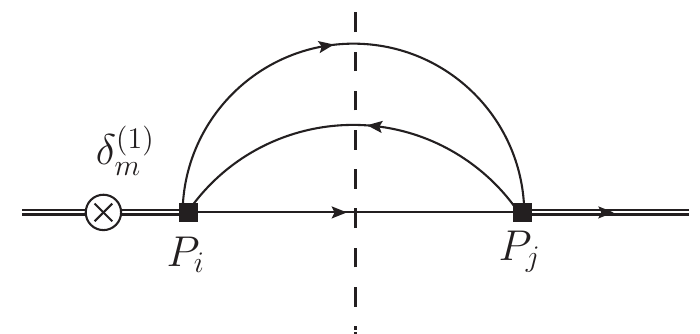}}
\end{minipage}
\\[1.2em]
%%%%%%%%%%%%%%%%%%%%%%%%%%%%%%%%%%%%%%%%%%%%%%%%%%%
\vspace{-0.75cm}
\begin{minipage}{.45\linewidth}
\centering
\hspace{+0.25cm}
\subfloat[{}]{\includegraphics[scale=.5]{figs/Tij.pdf}}
\end{minipage}
$\longleftrightarrow$
%%%%%%%%%%%%%%%%%%%%%%%%%%%%%%%%%%%%%%%%%%%%%%%%%%%%%
\begin{minipage}{.45\linewidth}
\captionsetup[subfigure]{labelformat=empty}
\centering
\vspace{0.75cm}
\hspace{-0.5cm}
\subfloat[{\Large +}]{\includegraphics[scale=.35]{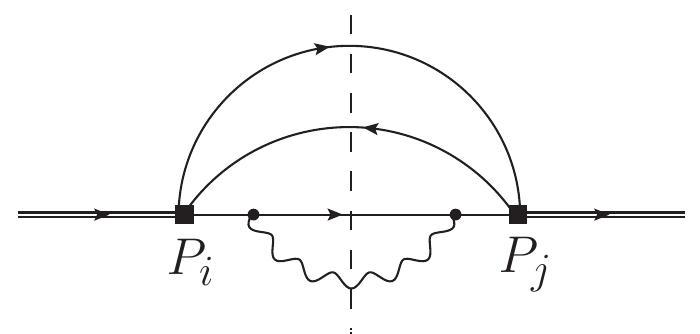}}
\hspace{0.5cm}
\subfloat[]{\includegraphics[scale=.35]{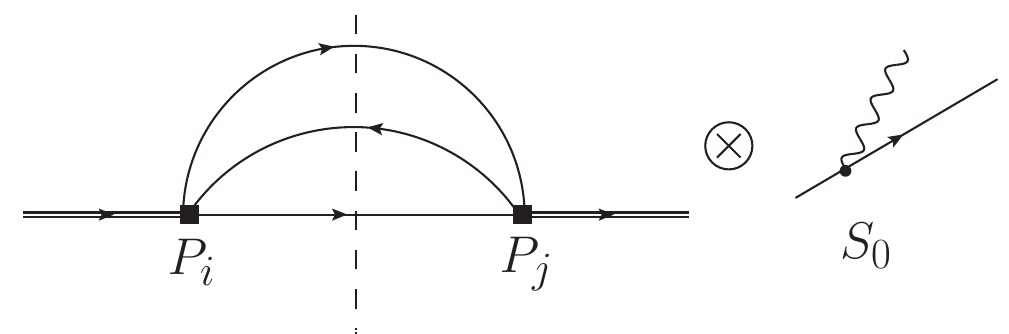}}
\end{minipage}
\caption{Sample diagrams of the building blocks from eq.~\eqref{eq:Masterformula}. Panel~(a) depicts a contribution to $T_{ij}$, panel~(b) to $V_{ij}$,~(c) to $R_{ij}$ and~(d) to $M_{ij}$. Photons can attach at all places possible and are not shown for clarity. For $T_{ij}$, $V_{ij}$, and $R_{ij}$ the symbols are understood to also contain the convolution of the diagrams without a photon with the LO subtraction kernel $S_0$, as shown in panel~(e). \label{fig:Masterformulaexample}}
\end{figure}

Turning now to $G_{ij}^{(1)}(\mu,\delta)$, it involves the computation of multiple building blocks which are sorted via the master formula 
\begin{align}
\label{eq:Masterformula}
    G_{ij}^{(1)}(\mu,\delta)=& \,V_{ji}+ V^{*}_{ij}+ R_{ji}+S_{ji}+ Z_{\psi} T_{ji} + Z_{\psi} T_{ij}^{*} + M_{ji} + M^{*}_{ij}+\sum_{k} \bigl\{Z_{jk} T_{ki} + Z_{ik} T_{kj}^{*}\bigr\},
\end{align}
where $V_{ij}$ are the one-loop four-body contributions of the $P_{i}$~--~$P_{j}$ interference where, by definition, the loop is to the left of the cut. $R_{ij}$ are the corresponding tree-level five-body contributions, $T_{ij}$ are the tree-level four-body contributions, and $M_{ij}$ are the mass-counterterm diagrams. Sample diagrams for these contributions can be seen in figure~\ref{fig:Masterformulaexample}. Furthermore, $Z_{\psi}$ are the wave-function counterterms  and $Z_{ij}$ the operator counterterms. Note that most of the building blocks of $G_{ij}^{(1)}(\mu,\delta)$ are not simply the bare amplitudes, as these can still contain IR divergences. In order to obtain an IR-finite result we employ the following IR-subtraction procedure,
\begin{alignat}{4}\label{eq:IRsubprocedure}
    T_{ij}&= T^{\text{bare}}_{ij}+T^{\slashed{\gamma_c}}_{ij} \otimes S_{0} \,,\nonumber && \quad 
R_{ij} && = R^{\text{bare}}_{ij}+R^{\slashed{\gamma_c}}_{ij} \otimes S_{0}\,,\\
V_{ij} & = V^{\text{bare}}_{ij}+V^{\slashed{\gamma_c}}_{ij} \otimes S_{0} \,, && \quad S_{ij} && = T^{\slashed{\gamma_c}}_{ij} \otimes S_{1} \, ,
\end{alignat}
where the label ``bare'' denotes the bare amplitude and the label ``$\slashed{\gamma_c}$'' is the associated diagram without additional photon radiation. Finally, $S_0$ and $S_1$ are the LO and NLO IR-subtraction kernels, which will serve to translate dimensionally into mass-regulated collinear divergences. Their explicit definitions will be given in section~\ref{sec:renormalisation}, and the symbol ``$\otimes$'' denotes a convolution. The sum of terms on the RHS of eq.~\eqref{eq:IRsubprocedure} is sketched in figure~\ref{fig:Masterformulaexample}(e).
%%%%%%%%%%%%%%%%%%%%%%%%%%%%%%%%%%%%%%%%%%%%%%%%%%%%%%%%
%%%%%%%%%%%%%%%%%%%%%%%%%%%%%%%%%%%%%%%%%%%%%%%%%%%%%%%

\section{Bare Calculation}
\label{Sec:barecalculation}

\subsection{Diagram generation, operator insertions, Dirac and colour algebra}
\label{Sec:processdiags}

In this chapter we cover the calculation of the one-loop four-body terms $V^{\text{bare}}_{ij}$ and the tree-level five-body terms $R^{\text{bare}}_{ij}$ from eq.~\eqref{eq:IRsubprocedure}. They amount to the parton-level processes
\begin{align}
b(p_b) & \, \to q(p_1) \, \bar q(p_2) \, s(p_3) \, \gamma(p_4) \, , \\[0.6em]
b(p_b) & \, \to q(p_1) \, \bar q(p_2) \, s(p_3) \, \gamma(p_4) \, g(p_5) \, ,
\end{align}
respectively, where $p_b^2=m_b^2$ and $p_i^2=0$, $i=1,\ldots,5$. To impose the aforementioned energy cut on the photon we define in the rest frame of the $b$ quark
\begin{equation}\label{eq:Egammapipj}
E_{\gamma} = \frac{p_b \cdot p_\gamma}{m_b} = \frac{p_b \cdot p_4}{m_b} 
\end{equation}
and parametrise the photon energy by the variable
\begin{equation}\label{eq:defz}
\frac{2 \, E_\gamma}{m_b} \equiv 1-z \equiv \bar z \, ,
\end{equation}
where we introduce the usual notation $\bar z = 1-z$, which we will also use for other variables in the following. Furthermore, the lower cut $E_0$ on the photon energy is parametrised via
\begin{equation}\label{eq:defdelta}
\frac{2 \, E_0}{m_b} \equiv 1-\delta \, ,
\end{equation}
such that the inequality $E_0 \le E_\gamma \le m_b/2$ is mapped onto $0 \le z \le \delta$, respectively $\bar\delta \le \bar z \le 1$.

\begin{figure}[t!]
\centering
\begin{minipage}{.45\linewidth}
\centering
\subfloat[{}]{\includegraphics[scale=.55]{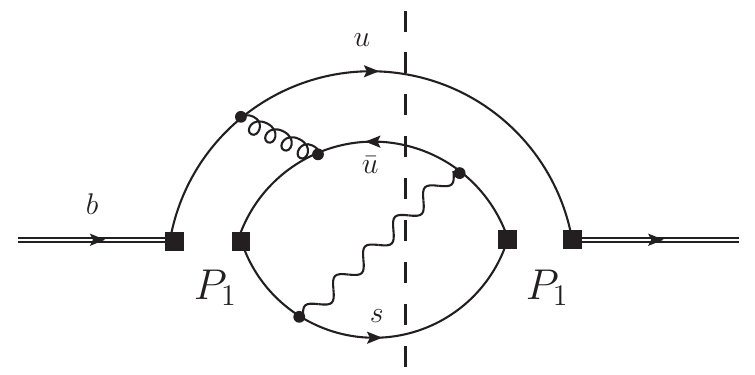}}
\end{minipage}
\begin{minipage}{.45\linewidth}
\vspace{-0.45cm}
\subfloat[{}]{\includegraphics[scale=.55]{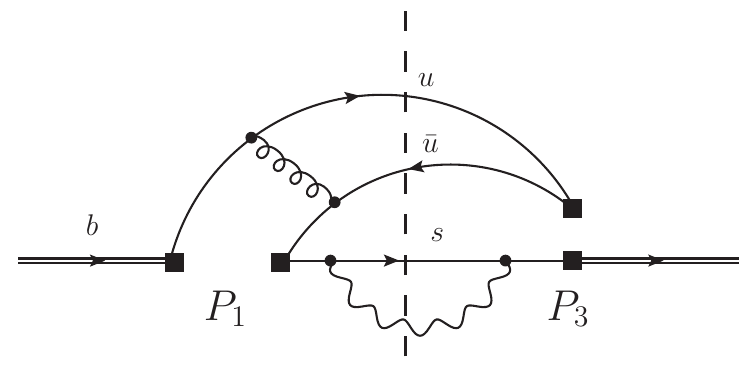}}
\end{minipage}
\\
\begin{minipage}{.45\linewidth}
\centering
\subfloat[{}]{\includegraphics[scale=.55]{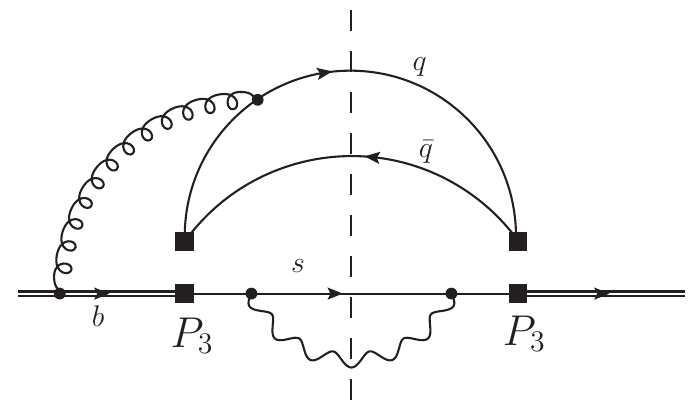}}
\end{minipage}
\begin{minipage}{.45\linewidth}
\centering
\subfloat[{}]{\includegraphics[scale=.55]{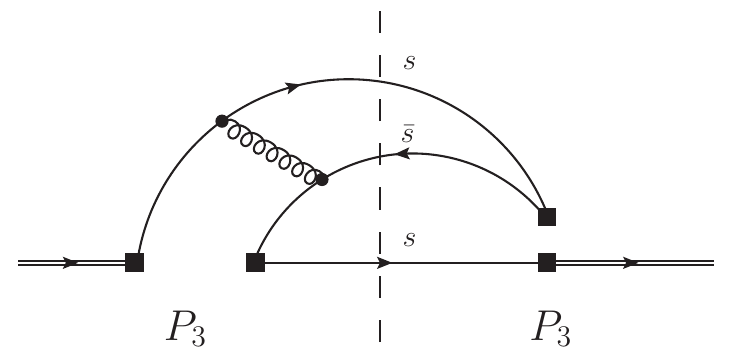}}
\end{minipage}
\caption{Sample diagrams for different operator insertions. Cases (a) and (b) only allow for $q=u$, while penguin-penguin insertions allow for $q=u,d,s$ in the final state  (cf.\ panel (c)). Penguin-penguin interferences with $q=s$ allow for an additional insertion (d).
\label{fig:examplediags4B}}
\end{figure}

We have to take all interferences of $\{P_{1,2}^u,P_{3-6}\}$ and therein various operator insertions into account, samples of which are depicted in figure~\ref{fig:examplediags4B}. The diagrams are generated with \texttt{QGRAF}~\cite{Nogueira:1991ex}, which yields 176 diagrams for each operator insertion for the four-body contribution, while for the five-body case we encounter 400 diagrams per insertion. The expressions are then passed on to \texttt{FORM}~\cite{Ruijl:2017dtg} and an in-house routine for performing the Dirac and colour algebra. Since the expressions contain traces with the Dirac matrix $\gamma_5$, special care is required in order to treat the latter consistently in dimensional regularisation with $D=4-2\epsilon$. We dedicate the next subsection to this issue.

%%%%%%%%%%%%%%%%%%%%%%%%%%%%%%%%
%%%%%%%%%%%%%%%%%%%%%%%%%%%%%%%%

\subsection{Treatment of \texorpdfstring{$\gamma_5$}{}}\label{sec:gamma5}

The Dirac matrix $\gamma_5$ is introduced in weak processes as part of the axial current $\gamma_{\mu}\gamma_5$. It is well-known that $\gamma_5$ is an inherently four-dimensional object, and that cyclicity of the trace and the anti-commutation relation
$ \{\gamma_5,\gamma_{\mu} \} = 0 $ cannot be maintained simultaneously in a consistent way when moving from four to $D$ dimensional space-time. A very comprehensive analysis of the ambiguities that can arise in calculations with weak axial currents is given in~\citere{Collins:1984xc}.

In order to treat the occurring expressions containing $\gamma_5$ consistently in $D$ dimensions, many schemes have been proposed, such as naive dimensional regularisation~\cite{Chanowitz:1979zu}, the 't Hooft-Veltmann scheme~\cite{tHooft:1972tcz}, and the Larin scheme~\cite{Larin:1993tq}. For our calculation, we adapted the so-called KKS scheme, which was developed by Kreimer, Körner and Schilcher~\cite{Korner:1991sx, Kreimer:1989ke}. It uses anti-commutativity $ \{\gamma_5,\gamma_{\mu} \} = 0 $ but not cyclicity of the trace. It is a so-called reading-point scheme and can be broken down to the following rules that one has to implement for traces containing $\gamma_5$.
\begin{itemize}
\item[i)] The anti-commutation relations $ \{\gamma_5,\gamma_{\mu} \} = 0$ and $\{\gamma_\mu,\gamma_{\nu} \} = 2g_{\mu\nu}$ hold.
\item[ii)] It is forbidden to use cyclicity in traces involving an odd number of $\gamma_5$
matrices.
\item[iii)] If applicable, start traces at an axial vertex $\gamma_{\mu} \gamma_5$. If there are several diagrams contributing to a given process, all traces must be read starting at the same vertex, called the {\emph{reading point}}. In the case
of several axial vertices, average over all possible axial starting points (`bosonisation'), e.g.
\begin{equation}
tr(\gamma_{\mu_2}\gamma_{5}\gamma_{\mu_3}\gamma_{\mu_4}\gamma_{5}\gamma_{\mu_1}) \to \frac{1}{2} \left( tr(\gamma_{\mu_2}\gamma_{5}\gamma_{\mu_3}\gamma_{\mu_4}\gamma_{5}\gamma_{\mu_1})+tr(\gamma_{\mu_4}\gamma_{5}\gamma_{\mu_1}\gamma_{\mu_2}\gamma_{5}\gamma_{\mu_3}) \right)\label{eq:bosonization}
\end{equation}
\item[iv)] Anticommute all occurring $\gamma_5$ matrices to the end of the trace. Use $\gamma_5^2=1$ until at most a single $\gamma_5$ remains. 
In the latter case, $\gamma_5$ is replaced by $\frac{i}{4!} \varepsilon^{\mu\nu\rho\sigma} \gamma_{\mu} \gamma_{\nu}\gamma_{\rho}\gamma_{\sigma}$.
\end{itemize}
As~\fig{fig:examplediags4B} shows, we encounter two different types of trace topologies. In the case of a single trace, we are either left with a trace free of $\gamma_5$, which can be treated in the usual way, or with a trace containing a single $\gamma_5$, resulting in a single $\varepsilon$-tensor. We will show below that the latter terms vanish after phase-space integration. In the case of a product of two Dirac traces, there is an additional case that we have to consider: If two traces that are multiplied contain a single $\gamma_5$ each, we end up with a product of two Levi-Civita tensors. In the CMM operator basis this case only appears in interferences of current-current operators $P_{1,2}^u$ with themselves, i.e.~in the upper left $2 \times 2$ block of $G_{ij}$. The product of two Levi-Civita tensors is linked to a determinant of metric tensors,
\begin{equation}
\varepsilon^{\mu_1 \mu_2 \mu_3 \mu_4} \varepsilon^{\nu_1 \nu_2 \nu_3 \nu_4}  =  \mathrm{det} \left(\begin{array}{ccc} 
g^{\mu_1 \nu_1} & \ldots & g^{\mu_1 \nu_4} \\ 
\vdots & \ddots & \vdots \\ 
g^{\mu_4 \nu_1} & \ldots & g^{\mu_4 \nu_4} \\ 
\end{array}  \right) \, ,
\end{equation}
whose implementation respects the symmetry properties and restores the correct dependence on the dimension. The resulting metric tensors are then handled as $D$-dimensional objects which is discussed, for example, in~\citere{Chen:2023lus} (see also~\cite{Chen:2024hlv}). Applying this scheme allowed us to replicate the tree-level results from~\citere{Kaminski:2012eb}.

%%%%%%%%%%%%%%%%%%%%%%%%%%%%%%%%%%%%%%%%%%%%%%%%%%%%%%

\subsection{Integral reduction}\label{Sec:IBP}

After these steps our expressions undergo an integral reduction based on integration-by-parts (IBP) identities~\cite{Chetyrkin:1981qh,Tkachov:1981wb} and Laporta's algorithm~\cite{Laporta:2000dsw}. To this end, all integrals are converted to four-loop propagator-type integrals via the method of reverse unitarity~\cite{Anastasiou:2002yz},
\begin{equation}\label{eq:cutkoskysimp}
-2 \pi i \, \delta(A) = \frac{1}{A+i\eta} - \frac{1}{A-i\eta}  \,,
\end{equation}
which trades phase-space constraints and the cut-condition of the photon energy (see eq.~\eqref{eq:cutrel} below) for propagators of loop integrals, and makes use of the fact that the integral reduction is inert to the $i\eta$ prescription. This method has the additional advantage of being insensitive to whether a four- or five-particle cut is taken in the end. We find that in this way, all expressions can be cast into one of the following seven integral topologies,
\begin{align}
{\cal{T}}_1 & = \Big\{ (p_b - p_{123})^2, 2p_b(p_b-p_{123})-m_b^2 \, \bar z,p_1^2,p_2^2,p_3^2,k^2,(p_2+k)^2,(p_3-k)^2, \nonumber \\[0.6em]
& \qquad (p_b-p_{12})^2, (p_b-p_{12}-k)^2,(p_1-k)^2,(p_b-p_{13})^2,(p_b-p_1)^2,(p_b-p_3)^2\Big\} \, ,  \\[1.0em]
{\cal{T}}_2 & = \Big\{ (p_b - p_{123})^2, 2p_b(p_b-p_{123})-m_b^2 \, \bar z,p_1^2,p_2^2,p_3^2,k^2,(p_3-k)^2,(p_{123}-k)^2-m_b^2, \nonumber \\[0.6em]
& \qquad (p_b-k)^2-m_b^2, (p_b-p_{12})^2,(p_1-k)^2,(p_b-p_{13})^2,(p_b-p_1)^2,(p_b-p_3)^2\Big\} \, ,  \\[1.0em]
{\cal{T}}_3 & = \Big\{ (p_b - p_{123})^2, 2p_b(p_b-p_{123})-m_b^2 \, \bar z,p_1^2,p_2^2,(p_b-p_{12}-k)^2,k^2,(p_3-k)^2,(p_{123}-k)^2-m_b^2, \nonumber \\[0.6em]
& \qquad (p_b-k)^2-m_b^2, (p_b-p_{12})^2,(p_1-k)^2,(p_b-p_{13})^2,(p_b-p_1)^2,(p_b-p_3)^2\Big\} \, ,  \\[1.0em]
{\cal{T}}_4 & = \Big\{ (p_b - p_{123})^2, 2p_b(p_b-p_{123})-m_b^2 \, \bar z,p_1^2,p_2^2,p_3^2,k^2,(p_b-p_{12}-k)^2,(p_{123}-k)^2-m_b^2, \nonumber \\[0.6em]
& \qquad (p_b-k)^2-m_b^2, (p_b-p_{12})^2,(p_1-k)^2,(p_b-p_{13})^2,(p_b-p_1)^2,(p_b-p_3)^2\Big\} \, ,  \\[1.0em]
{\cal{T}}_5 & = \Big\{ (p_b - p_{123})^2, 2p_b(p_b-p_{123})-m_b^2 \, \bar z,p_1^2,p_2^2,p_3^2,k^2,(p_3-k)^2,(p_b-p_{12}-k)^2, \nonumber \\[0.6em]
& \qquad (p_b-k)^2-m_b^2, (p_b-p_{12})^2,(p_1-k)^2,(p_b-p_{13})^2,(p_b-p_1)^2,(p_b-p_3)^2\Big\} \, ,  \\[1.0em]
{\cal{T}}_6 & = \Big\{ (p_b - p_{123})^2, 2p_b(p_b-p_{123})-m_b^2 \, \bar z,p_1^2,p_2^2,p_3^2,k^2,(p_2+k)^2,(p_3-k)^2, \nonumber \\[0.6em]
& \qquad (p_b-p_{12})^2, (p_b-p_{12}-k)^2,(p_1+k)^2,(p_b-p_{13})^2,(p_b-p_1)^2,(p_b-p_3)^2\Big\} \, ,  \\[1.0em]
{\cal{T}}_7 & = \Big\{ (p_b - p_{123})^2, 2p_b(p_b-p_{123})-m_b^2 \, \bar z,p_1^2,p_2^2,p_3^2,k^2,(p_2+k)^2,(p_b-p_{13}-k)^2, \nonumber \\[0.6em]
& \qquad (p_b-p_{12})^2, (p_b-p_{12}-k)^2,(p_1-k)^2,(p_b-p_{13})^2,(p_b-p_1)^2,(p_b-p_3)^2\Big\} \, .
\end{align}

\begin{figure}[t!]
\centering
\includegraphics[width=10.8cm]{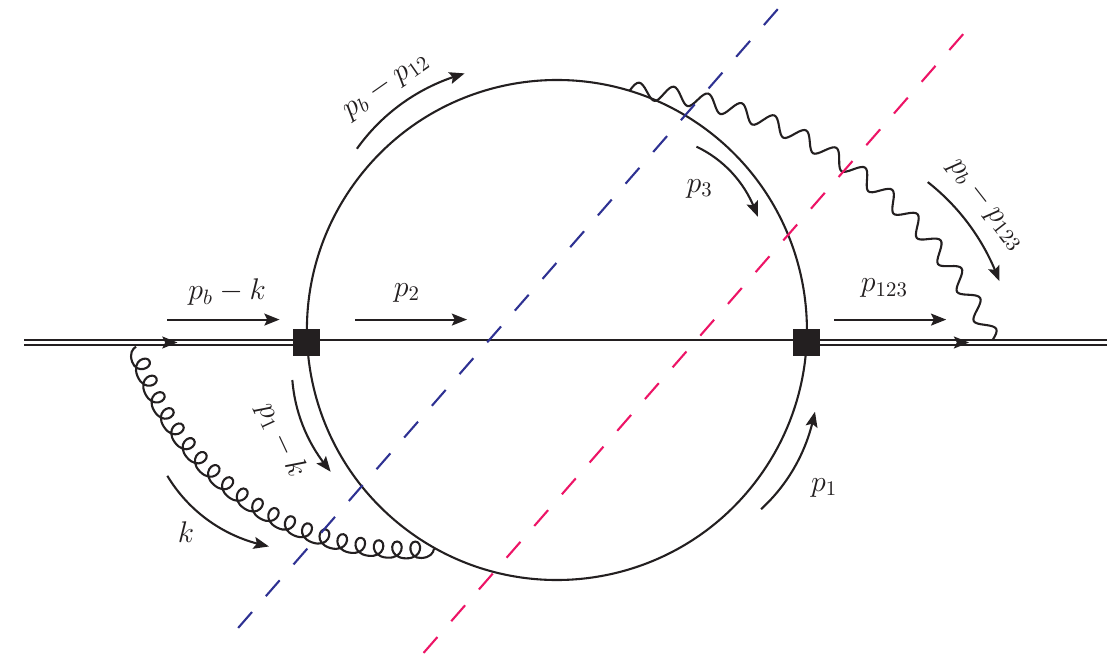}
\caption{A sample diagram which fits in topology ${\cal{T}}_2$, together with a four (red) and five (blue) particle cut. As usual, wavy (curly) lines denote photons (gluons), whereas single (double) solid lines stand for massless (massive) quarks. The black squares denote operator insertions from the effective Hamiltonian.\label{fig:exampleT2}}
\end{figure}

\begin{figure}[htbp]
\includegraphics[width=0.84\linewidth]{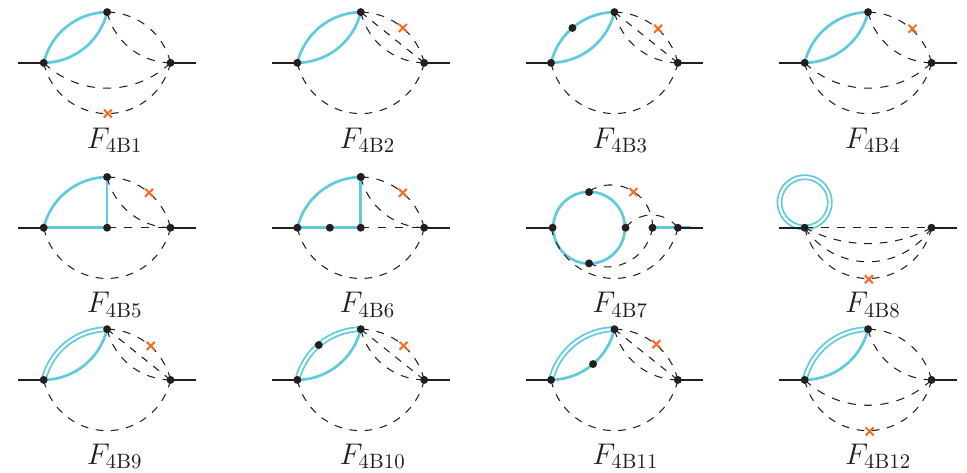}\\
\includegraphics[width=0.84\linewidth]{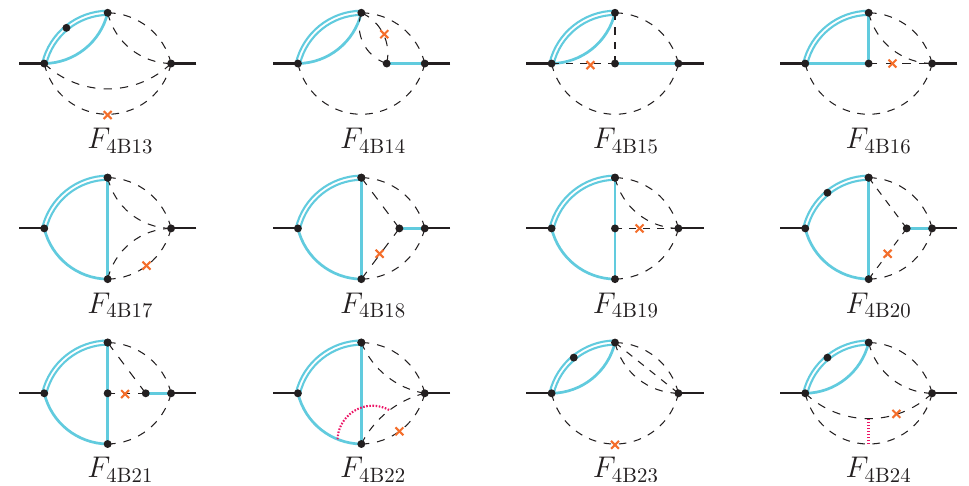}\\
\includegraphics[width=0.84\linewidth]{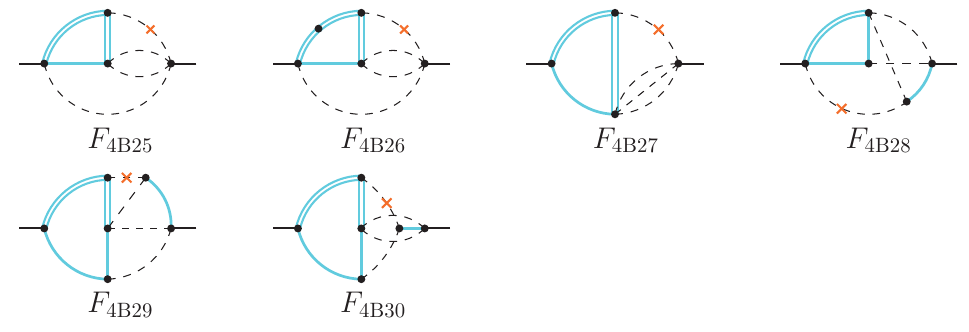}
\vspace*{-3pt}
\caption{One-loop four-body master integrals. The dashed lines indicate the cut propagators, the solid light blue single (double) lines indicate massless (massive) propagators. A dot on a line indicates a squared propagator. The dotted red lines connecting lines with momenta $l_1$ and $l_2$ denote numerators $(l_1-l_2)^2-m_1^2-m_2^2$. The orange cross denotes the cut photon propagator with momentum $p_4$. Due to the energy cut, this line breaks the symmetry in the final-state momenta.\label{fig:4BMIs}}
\end{figure}

\begin{figure}[htbp]
\begin{center}
\includegraphics[width=0.84\linewidth]{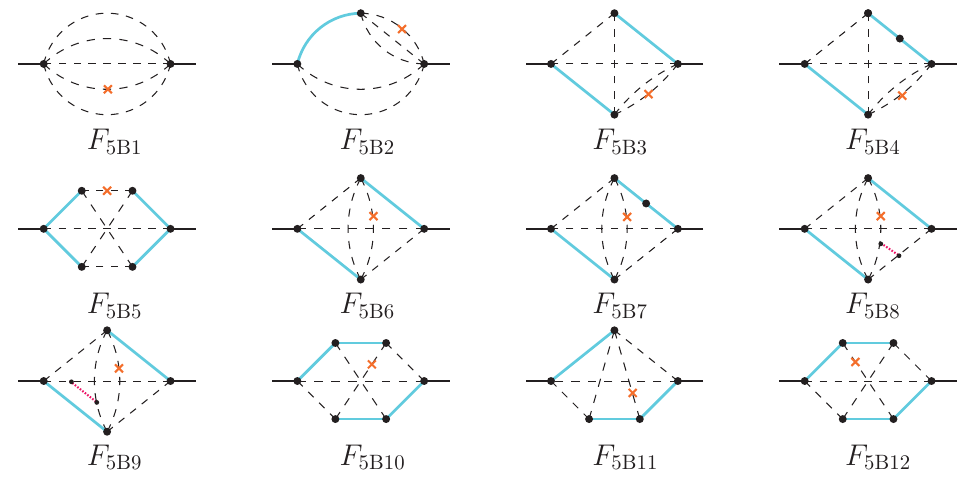}
\includegraphics[width=0.84\linewidth]{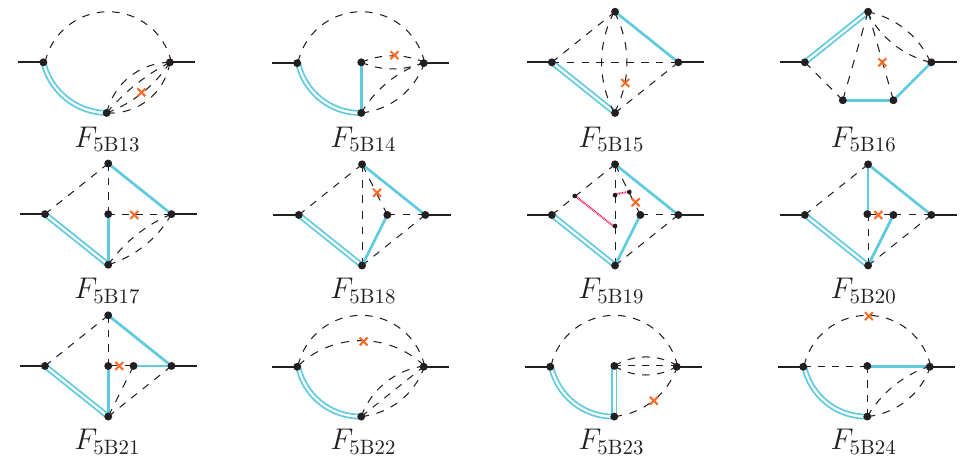}
\includegraphics[width=0.84\linewidth]{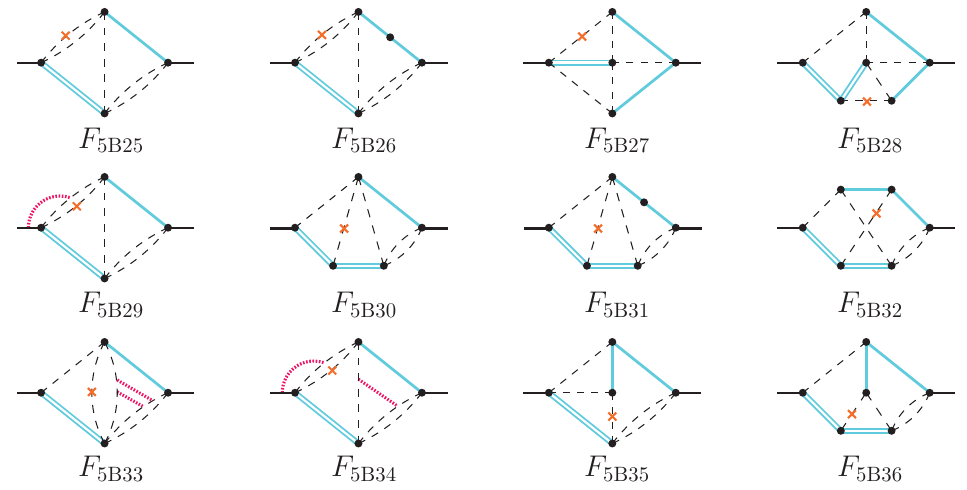}
\end{center}
\caption{Master integrals from the five-body bremsstrahlung contribution. Symbols have the same meaning as in figure~\ref{fig:4BMIs}.\label{fig:5BMIs}}
\end{figure}

Here $p_b$ with $p_b^2=m_b^2$ is the external and $p_1$, $p_2$, $p_3$, $k$ are the loop momenta. Moreover, we use the notation $p_{i\ldots j}=p_i + \ldots + p_j$. A sample diagram which fits into ${\cal{T}}_2$ is shown in figure~\ref{fig:exampleT2}. The integral reduction itself was performed in \texttt{FIRE}~\cite{Smirnov:2019qkx} on the local computing cluster \texttt{OMNI}. The runtime ranges from ${\cal{O}}({\text{hours}})$ to ${\cal{O}}({\text{days}})$ and the memory consumption from ${\cal{O}}({\text{MB}})$ to ${\cal{O}}({100~\text{GB}})$, depending on the number of integrals that has to be reduced in a certain topology and whether or not a topology contains massive propagators. When re-introducing the photon-energy and phase-space cuts after the reduction, any integral which contains a propagator that gets cancelled by a numerator (i.e.\ a line having a non-positive index) and subsequently cut, is set to zero due to the relation
\begin{equation}
    (p^2-m^2)^n \delta(p^2-m^2) = 0 \, , \qquad n = 1,2,\ldots \, .
\end{equation}
The master integrals that we find through the reduction procedure are depicted in figures~\ref{fig:4BMIs} and~\ref{fig:5BMIs} for the four and five-body cases, respectively. We note here in passing that the set of master integrals in these figures is slightly larger than what comes out of the reduction algorithm, but turns out to be convenient for setting up the system of differential equations (to be discussed in section~\ref{sec:MI}). The relations among the integrals are relegated to appendix~\ref{app:integralrelations} and will serve as a cross-check for the results of master integrals.

%%%%%%%%%%%%%%%%%%%%%%%%%%%%%%%%%%%%%%%%%%%%%%%%%%%%

\afterpage{\FloatBarrier}
\afterpage{\FloatBarrier}
%%%%%%%%%%%%%%%%%%%%%%%%%%%%%%%%%%%%%%%%%%%%%%%%%%%%%%%%

\subsection{Phase space parametrisations}\label{sec:phasespacegeneral}

The master integrals have to be integrated over a multi-particle phase space. The dimensionally regulated phase space measure for a $1 \rightarrow n$ decay process of a $b$ quark is defined as
\begin{equation}\label{eq:PSgeneral}
dPS_n = \left( \prod_{f=1}^{n} \frac{d^{D-1} p_f}{(2\pi)^{D-1}} \frac{1}{2E_f} \right) (2 \pi)^D \, \delta^{(D)}(p_b-\sum_{f=1}^{n} p_f) \,,
\end{equation}
where $p_b$ is the momentum of the incoming $b$ quark and the $p_f$ ($E_f$) are the momenta (energies) of the particles in the final state, which in our calculation are taken to be massless. For our purposes $n=3$, $4$, and $5$ are required. The above formula is often rewritten in terms of angles and the dimensionless invariants $s_{ij}=2\, p_i \cdot p_j/m_b^2$. In our case this procedure trivialises the integrations over the angular volumes using
\begin{equation}
V(D)=\int d \Omega_D = \frac{2 \pi^{D/2}}{\Gamma(\frac{D}{2})}\, ,
\end{equation}
and renders only the dependence on the $s_{ij}$ non-trivial. After this change of variables, one arrives at the following expressions for the different number of particles (see e.g.~\citeres{Gehrmann-DeRidder:2003pne,Heinrich:2006sw}),
\begin{align}
dPS_2 & = (2\pi)^{2-D} (m_b^2)^{\frac{D-4}{2}}  (s_{12})^{\frac{D-4}{2}} \frac{d \Omega_{D-1}}{2^{D-1}} \ ds_{12} \ \delta(1-s_{12}) \,, \\[0.8em]
dPS_3 & = (2\pi)^{3-2D} \, 2^{-1-D} (m_b^2)^{D-3} (s_{12}s_{13}s_{23})^{\frac{D-4}{2}}\ \delta(1-s_{12}-s_{13}-s_{23})\\ \nonumber
		& \quad \;\, \times d\Omega_{D-1}\ d\Omega_{D-2} \ ds_{12} \ ds_{13}\ ds_{23} \,, \\[0.8em]
dPS_4 & = (2\pi)^{4-3D} \, 2^{1-2D} (m_b^2)^{\frac{3D-8}{2}} \delta(1-s_{12}-s_{13}-s_{23}-s_{14}-s_{24}-s_{34})\\ \nonumber
	   & \quad \;\, \times (-\Delta_4)^{\frac{D-5}{2}}\Theta(-\Delta_4)\ d\Omega_{D-1}\ d\Omega_{D-2}\ d\Omega_{D-3} \ ds_{12} \ ds_{13}\ ds_{23} \ ds_{14} \ ds_{24}\ ds_{34} \, , \\[0.8em]
dPS_5 & = (2\pi)^{5-4D} \, 2^{-2-2D} \, (m_b^2)^{2D-5} \, d\Omega_{D-1}\ d\Omega_{D-2}\ d\Omega_{D-3}\ d\Omega_{D-4}  \nonumber \\
& \quad \;\, \times \delta(1-s_{12}-s_{13}-s_{23}-s_{14}-s_{24}-s_{34}-s_{15}-s_{25}-s_{35}-s_{45}) \nonumber \\
& \quad \;\, \times (-\Delta_5)^{\frac{D}{2}-3} \Theta(-\Delta_5) \ ds_{12} \ ds_{13}\ ds_{23} \ ds_{14} \ ds_{24}\ ds_{34} \ ds_{15} \ ds_{25}\ ds_{35} \ ds_{45}\,.
\end{align}
$\displaystyle \Delta_n$ is related to the determinant of the Gram matrix $G_n$ with entries $G_{ij} = 2 p_i\cdot p_j$, $i,j=1,\ldots,n$. One has
\begin{equation}
\Delta_4 = \frac{1}{(m_b^2)^4} \, \mathrm{det}(G_4) = \lambda(s_{12}s_{34},s_{13}s_{24},s_{14}s_{23})
\end{equation} 
with the K\"all\'en function $\lambda(x,y,z)= x^2+y^2+z^2 -2 (xy+xz+yz)$, and
\begin{align}
\Delta_5 &= -\frac{1}{2 (m_b^2)^5} \, \mathrm{det}(G_5) = + y_{10}^2 y_1 y_{2} y_{3} + y_{9}^2 y_{1} y_{4} y_{5} + y_{8}^2 y_{2} y_{4} y_{6} + y_{7}^2 y_{3} y_{5} y_{6} + y_{6}^2 y_{1} y_{7} y_{8}  \nonumber \\
& + y_{5}^2 y_{2} y_{7} y_{9} + y_{4}^2 y_{3} y_{8} y_{9} + y_{3}^2 y_{4} y_{7} y_{10} + y_{2}^2 y_{5} y_{8} y_{10}+ y_{1}^2 y_{6} y_{9} y_{10} \nonumber \\
&- y_{10} [ y_{2} y_{3} y_{5} y_{7} + y_{1} y_{3} y_{6} y_{7} + y_{2} y_{3} y_{4} y_{8} + y_{1} y_{2} y_{6} y_{8} + y_{1} y_{3} y_{4} y_{9} + y_{1} y_{2} y_{5} y_{9}] \nonumber \\
&-y_{9} [ y_{4} y_{5}(y_{3} y_{7} + y_{2} y_{8}) + y_{1} y_{6} (y_{5} y_{7} + y_{4} y_{8})] - y_{6} y_{7} y_{8} (y_{3} y_{4} + y_{2} y_{5}) \, ,
\label{eq:5bodyPara}
\end{align} 
expressed in terms of the variables
\begin{align}
y_1 &= s_{12} \, , && y_2 = s_{13}  \, , && y_3 = s_{23} \, , && y_4 = s_{14} \, , && y_5 = s_{24} \nonumber \\
y_6 &= s_{34} \, , && y_7 = s_{15}  \, , && y_8 = s_{25} \, , && y_9 = s_{35}\, , && y_{10} = s_{45} \, .
\end{align}

\subsection{Phase space integration}
\label{sec:4particlePS}

The implementation of the photon-energy cut into the phase-space integrations is then done by introducing an additional integration over $z$ and a delta function, a method that was for example used in~\citere{Huber:2014nna}. We exemplify this procedure via the four-body decay
\begin{equation}
b (p_b) \rightarrow q(p_1) \, \bar{q}(p_2) \, s(p_3) \, \gamma(p_4) \, ,
\end{equation}
in which case $\bar z = 2 \, E_{\gamma}/m_b = s_{14}+s_{24}+s_{34}$. We multiply the phase-space integrand by the function $\delta(\bar z-s_{14}-s_{24}-s_{34})$ and introduce an integration over $z$ from 0 to $\delta$. The relevant phase-space integrations then read
\begin{equation}\label{eq:cutrel} 
\int_0^{\delta} dz \int_0^1 [ds_{ij}] \ \delta(\bar z-s_{14}-s_{24}-s_{34}) \delta(z-s_{12}-s_{13}-s_{23}) \ (-\Delta_4)^{\frac{D-5}{2}} \Theta(-\Delta_4) \ \mathcal{K}(s_{ij}) \,,
\end{equation}
where $\mathcal{K}$ is the integration kernel, the first delta function fixes $E_\gamma$ and the other one originates from momentum conservation. Using these to fix two of the invariants, we arrive at five integrations left to perform,
\begin{align}
\Gamma_{E_{\gamma}>E_0} =N(D) \int_0^{\delta} dz \int_0^{\bar{z}} ds_{34}   \int_0^{\bar{z}-s_{34}} ds_{14} & \int_0^z ds_{12}   \int_0^{z-s_{12}} ds_{23} \,  \nonumber   \\
& \times \,\mathcal{K}(s_{ij}) \ (-\Delta_4)^{\frac{D-5}{2}} \Theta(-\Delta_4) \ \Bigg|_{\substack{s_{13} = z-s_{12}-s_{23}\\s_{24} = \bar{z}-s_{14}-s_{34}}} \, .
\end{align}
In this equation we also introduced the normalisation factor
\begin{equation}
\displaystyle N(D) =  \frac{\tilde{\mu}^{6\epsilon} 2^{8-5D} \pi^{1-3D/2} m_b^{3D-9}}{4 N_c \Gamma(\frac{D-1}{2})\Gamma(\frac{D-2}{2})\Gamma(\frac{D-3}{2})}  \, ,
\end{equation}
such that from the general formula for the unpolarised decay rate
\begin{equation}
\Gamma_{E_{\gamma}>E_0} = \frac{1}{2 m_b} \frac{1}{2 N_c} \int_{E_{\gamma}>E_0}  dPS_4  \ \sum_{\substack{\mathrm{spin}\\ \mathrm{colour}}} | \mathcal{M} |^2
\end{equation}
we deduce  $\mathcal{K}(s_{ij}) = \sum | \mathcal{M} |^2$. For practical purposes, it is also important to factorise $\Delta_4$ and to map the integrations onto the unit hypercube. We adapt here the variable transformation introduced in~\citeres{Gehrmann-DeRidder:2003pne,Huber:2014nna},
\begin{align}\label{eq:parametrization4b}
s_{12}&= z v w  &s_{34}&= \bar{z} \bar{v} \nonumber\\
s_{14}&= \bar{z} v x  & s_{23}&= (a^{+}-a^{-})u + a^{-} \nonumber\\
s_{13}&= z - s_{12} - s_{23} &s_{24}&=\bar{z}-s_{14}-s_{34}  \,,
\end{align}
with 
\begin{equation}
 a^{\pm} = z [\bar{v}wx+\bar{w}\bar{x} \pm 2(\bar{v}w\bar{w}x\bar{x})^{1/2}] \,.
\end{equation}
With this, the phase-space measure now completely factorises,
\begin{equation}
\Gamma_{E_{\gamma}>E_0} = N(D)\ 4^{D-4} \int_0^{\delta} dz\, (z \bar{z})^{D-3} \int_0^1 du\, dv\, dx\, dw\, (u\bar{u})^{\frac{D-5}{2}} v^{D-3} (\bar{v}x \bar{x}w \bar{w})^{\frac{D-4}{2}} \, \mathcal{K} \,.
\end{equation}

The five-body case
\begin{equation}
b (p_b) \rightarrow q(p_1) \, \bar{q}(p_2) \, s(p_3) \, \gamma(p_4) \, g(p_5) \, ,
\end{equation}
is treated in a similar manner, now having $\bar z = s_{14}+s_{24}+s_{34}+s_{45}$. The 
factorisation of the phase-space measure and mapping onto the unit hypercube is achieved by means of the following transformation~\cite{Heinrich:2006sw},
\begin{alignat}{3}
s_{1345} &= t_7 \,,&& \quad s_{34} && = t_2 t_6 t_7 \bar{t}_4  \,, \nonumber\\
s_{134} &= t_6 t_7\,, && \quad s_{15}+s_{45} &&= t_7 \bar t_6 [1-t_9(1-t_2 t_4)]\,, \nonumber\\
s_{13} &= t_6 t_7 \bar{t}_2 \,, && \quad  s_{25} &&= y_8^- +(y_8^+ - y_8^-)t_8 \, ,\nonumber\\
s_{23} &= t_3 \bar{t}_7 (1-t_2 t_4)(t_6 \bar{t}_9+t_9)\,, && \quad s_{35} &&= t_7 t_9 \bar{t}_6 (1-t_2 t_4)\,,\nonumber\\
s_{14} &= t_2 t_4 t_6 t_7 \, ,&& \quad s_{45} &&= y_{10}^- + (y_{10}^+ - y_{10}^-) t_{10}\,,\nonumber\\
s_{24} &= y_5^- + (y_5^+-y_5^-)t_5\, , &&  && \label{eq:factorizedPS5}
\end{alignat}
where $s_{12}$ (respectively $y_1$) has been eliminated by momentum conservation, and we use $s_{ijk} = s_{ij} + s_{ik} +s_{jk}$ and $s_{ijkl} = s_{ij} + s_{ik} +s_{jk} +s_{il} + s_{jl} +s_{kl} $ for the triple respectively quadruple invariants. The theta-function constraint $\Theta(-\Delta_5)$ is solved for $y_5$, which gives the solutions
\begin{equation}
y_5^{\pm} = y_5^0 \pm \sqrt{R_5} \,,
\end{equation}
then $\sqrt{R_5} \geq 0$ is solved for $y_8$ and finally  $y_8^+-y_8^- \geq 0$ is solved for $y_{10}$. While the expressions for $y_5^{\pm}$ are too lengthy to be given here (yet straightforward to derive), the expressions for $y_8^\pm$ and $y_{10}^\pm$ are
\begin{eqnarray}
y_8^\pm&=&y_8^0\pm d_8/2 \, ,\nonumber\\[0.3em]
y_8^0&=& \bar t_6 \, \bar t_7\,\{t_9 + t_3\,[t_6\,\bar t_9 - t_9]\}/
     (t_6\, \bar t_9 + t_9)\, ,\nonumber\\[0.3em]
d_8&=&y_8^+ -y_8^-=4\,\bar t_6 \, \bar t_7\,
\sqrt{\bar t_3 \, t_3\,t_6\,\bar t_9\,t_9}/
        (t_6\, \bar t_9 + t_9)\, ,\nonumber\\[0.3em]
y_{10}^\pm&=&y_{10}^0\pm d_{10}/2\, ,\nonumber\\[0.3em]
y_{10}^0&=&t_2\,t_7\, \bar t_6\,
\{ \bar t_9 - t_4\,[1 -t_9 (2 - t_2)\,]\}/
      (1 - t_2\,t_4)\, ,\nonumber\\[0.3em]
d_{10}&=&y_{10}^+-y_{10}^-=4\,t_7\,t_2\, \bar t_6\,
\sqrt{\bar t_2 \, \bar t_4 \,t_4\, \bar t_9\,t_9}/(1 - t_2\,t_4)\, .
\end{eqnarray}
As mentioned above, all integrations over $t_2,\ldots,t_{10}$ run from $0 \ldots 1$, and the Jacobian of the transformation~\eqref{eq:factorizedPS5} is
\begin{equation}
\mathrm{det}\left(\frac{\partial s_{ij}}{\partial t_k}\right) = 256 \sqrt{t_8 \, \bar t_8 \, t_{10} \, \bar t_{10}} \, t_2^3 \, \bar t_2 \, t_3 \, \bar t_3 \, t_4 \, \bar t_4 \, t_6^3  \, \bar t_6^{ \, 3} \, t_7^5 \, \bar t_7^{\, 3} \, t_9 \, \bar t_9 \, .
\end{equation}

%%%%%%%%%%%%%%%%%%%%%%%%%%%%%%%%%%%%%%%%%%%%%%%%%%%%%%

\subsection{Cancellation of terms proportional to \texorpdfstring{$\varepsilon^{\mu_1 \mu_2 \mu_3 \mu_4}$}{}}\label{sec:canceleps}

After going through the procedure of section~\ref{sec:gamma5}, we are left with expressions that contain at most a single $\varepsilon^{\mu_1 \mu_2 \mu_3 \mu_4}$. Similar to the arguments given in~\citere{Huber:2014nna} we show in the following that all pieces proportional to the antisymmetric $\varepsilon$-tensor cancel out after fully carrying out the angular integrations occurring in the phase space. To see this, let us consider the expression $\varepsilon_{\mu_1 \mu_2 \mu_3 \mu_4} p_1^{\mu_1}p_2^{\mu_2}p_3^{\mu_3}p_4^{\mu_4}$ in the rest frame of the $b$ quark and fix all momentum invariants $p_i \cdot p_j$. We can derive from momentum conservation that all the energies $p_b \cdot p_i$ are fixed. From this, we can also infer that the $\vec{p}_i \cdot \vec{p}_j $ are fixed, leaving us with the freedom to align the coordinate system. If we define a plane spanned by $\vec{p}_1$ and $\vec{p}_2$ this fixes $\vec{p}_3$ up to its orientation relative to that plane. Choosing this sign then also fixes $\vec{p}_4$. With this argument we can see that the expression $\varepsilon_{\mu_1 \mu_2 \mu_3 \mu_4} p_1^{\mu_1}p_2^{\mu_2}p_3^{\mu_3}p_4^{\mu_4}$ is fixed by the invariants $p_i \cdot p_j$ up to the aforementioned sign. The terms involving an $\varepsilon$-tensor that undergo phase-space integration are all of the type $F(p_i \cdot p_j)\ \varepsilon_{\mu_1 \mu_2 \mu_3 \mu_4} p_1^{\mu_1}p_2^{\mu_2}p_3^{\mu_3}p_4^{\mu_4}$. Since the function $F(p_i \cdot p_j)$ is parity-even (i.e.\ it does not depend on the orientation of the momenta), but $\varepsilon_{\mu_1 \mu_2 \mu_3 \mu_4} p_1^{\mu_1}p_2^{\mu_2}p_3^{\mu_3}p_4^{\mu_4}$ is parity-odd (as it changes sign under change of orientation of e.g.\ $\vec{p}_3$), this combination vanishes after integrating over the symmetric angular variables. We emphasise that this logic also holds true when imposing a cut on the photon energy since the cut-condition is also of the form $F(p_i \cdot p_j)$, see~\eqs{eq:Egammapipj} and~\eqref{eq:cutrel}.

%%%%%%%%%%%%%%%%%%%%%%%%%%%%%%%%%%%%%%%%%%%%%%%%%%%%%%%%%%%%%%%%%%%

\subsection{Evaluating master integrals}
\label{sec:MI}

The rational pre-factors that are generated during the integral reduction depend on the dimensional regulator $\eps$ and the dimensionless quantity $z$. Hence, the integration over the latter variable can only be carried out after the master integrals are substituted into the squared amplitude. We therefore evaluate all master integrals as a function of $z$ and $\eps$, $F_i = F_i(z,\eps)$. The simplest integrals can be evaluated by explicit phase-space integration, the more complicated ones by the method of differential equations.

\subsubsection{Differential equations}

For a given topology $i = 1,\ldots,7$, our master integrals obey a first-order, inhomogeneous, linear differential equation (DE)~\cite{Kotikov:1990kg,Remiddi:1997ny,Argeri:2007up} of the form
\begin{equation}
\partial_z \vec{F}_i = \hat{A}_i(z,\epsilon) \vec{F}_i \,,
\end{equation}
The matrices $\hat{A}_i(z,\epsilon)$ can now be brought into $\epsilon$-form~\cite{Henn:2013pwa}. There are multiple implementations of the algorithmic approach by Lee~\cite{Lee:2014ioa}, for our calculation we use the program \texttt{epsilon}~\cite{Prausa:2017ltv}. Running the program gives us the necessary ingredients to change our basis, namely the transformation matrices $T_i$ and the differential-equation matrices $A_{i,\epsilon}$. Changing our basis from $\vec{F}_i$ to $\vec{G}_i$, these matrices are appear in the following way,
\begin{align}
A_{i}(z,\epsilon) & = T_i A_{i,\epsilon}T_i^{-1}+(\partial_z T_i)T_i^{-1}\,, \\[0.3em]
\vec{G}_i & = \hat{T}^{-1}_i \vec{F}_i \,, \label{eq:trafotoUT}
\end{align}
and result in a new system of equations for each family,
\begin{equation}
\partial_z \vec{G}_i = \epsilon \hat{A}_{i,\epsilon}(z) \vec{G}_i \, ,
\end{equation}
where the dependence on the space-time is factorised from that on the kinematics. The structure of the matrices $\hat{A}_{i,\epsilon}(z)$ is such that the solution to the DE falls into the classes of harmonic polylogarithms (HPLs)~\cite{Remiddi:1999ew} and Goncharov polylogarithms~\cite{Goncharov:2001zfh}, defined, respectively, by
\begin{align}\label{eq:defHPL}
H_{\vec0_n}(x) & = \frac{1}{n!} \ln^n (x) \, ,\qquad\qquad
H_{a_1 , a_2 , \ldots, a_n }(x)  = \int_0^x dt \; f_{a_1}(t) \, H_{a_2, \ldots, a_n }(t)  \, ,
\end{align}
with weight functions $f_{0}(t) = 1/t$, $f_{\pm 1} = 1/(1\mp t)$, and
\begin{align}\label{eq:gpldef}
G_{\vec0_n}(x) & = \frac{1}{n!} \ln^n (x) \, , \qquad\qquad
G_{a_1,\ldots,a_n}(x)  = \int_0^x dt \ \frac{1}{t-a_1} \ G_{a_2,\ldots,a_n}(t) \,.
\end{align}
The weights that appear in our calculation are from the set $\displaystyle\{0,\pm 1,\pm i/\sqrt{3}\}$. To achieve this form for all topologies, some of the DE have to be formulated in the variables $\bar z$ or $x \equiv i \sqrt{\bar z/(4-\bar z)}$. The only missing ingredient for formulating the full solution are values at certain boundaries, usually $z=0$ or $z=1$. The methodology of finding the solutions for these (beyond trivial low-line integrals that we could obtain in fully analytic form) will be discussed next.

\subsubsection{Boundary conditions}

Here we give an example for each of the applied methods.

\subsection*{Case 1: Asymptotic behaviour with hypergeometric functions}
We consider the integral $F_{4\mathrm{B}9}$,
\begin{equation}
F_{4\mathrm{B}9} = \int dPS_4 \int \frac{d^D k}{(2\pi)^D} \frac{1}{{[{k}^2]} [{(k+p_{124})}^2-m_b^2]} \, ,
\end{equation}
where we suppress an overall factor of $(m_b^2)^{2D-6}$ in what follows to make the integral dimensionless, and tacitly assume that we stay differential in $z$. We use the parametrisation of the phase-space from section~\ref{sec:4particlePS} and introduce Feynman parameters. After carrying out the first integrations, this leads us to ($S_\Gamma = 1/((4\pi)^{D/2}\Gamma(1-\eps))$)
\begingroup
\allowdisplaybreaks
\begin{align}\label{eq:f4b9bc1}
F_{4\mathrm{B}9} &=  2\pi i S_{\Gamma}^4 \, \frac{ 2 \, \Gamma^6(1-\epsilon) \Gamma(\epsilon)}{\Gamma(3-2\epsilon)\Gamma(3-3\epsilon)}   \ (z\bar{z})^{1-2\epsilon} \nonumber \\
&\phantom{=}\, \times \int_0^1  dw \ (w\bar{w})^{-\epsilon} {}_3F_2(\epsilon,1,2-2\epsilon;2-\epsilon,3-3\epsilon;1-z\bar{w})\,,
\end{align}
\endgroup
which has no closed-form solution. However, as we only need the asymptotic behaviour in one point of $z$, we can set it to a definite value in the argument of the ${}_3F_2$ function and compute the expansion in $\epsilon$. The choice $z \to 1$ leaves the expansion in $\eps$ and integration over $w$ interchangeable and enables carrying out the $w$-integration in terms of another hypergeometric function, 
\begin{align}
F_{4\mathrm{B}9}|_{z\rightarrow 1} &= 2\pi i S_{\Gamma}^4 \ \frac{ 2 \, \Gamma^8(1-\epsilon) \, \Gamma(\epsilon)}{\Gamma(2-2\epsilon)\Gamma(3-2\epsilon)\Gamma(3-3\epsilon)} \ (z\bar{z})^{1-2\epsilon} \; {}_3F_2(\epsilon,1,1-\epsilon;2-\epsilon,3-3\epsilon;1) \nonumber \\[0.3em]
&=z \bar{z} \, \frac{i \pi S_{\Gamma}^4}{\epsilon} \Bigg[ 1+\eps \Big(-2 \log (\bar{z})-\frac{\pi ^2}{3}+13\Big) +\epssq \Big(2 \log ^2(\bar{z})  +\frac{2}{3} \left(\pi ^2-39\right) \log (\bar{z})\nonumber \\
& \qquad \qquad \quad -16 \zeta(3)-\frac{10 \pi ^2}{3}+103\Big) + \qfto(\epstrip) \Bigg] \,.
\end{align}
The expansion is done with {\texttt{HypExp}}~\cite{Huber:2005yg,Huber:2007dx}. The overall factor of $z\bar{z}$ has to be kept explicit in this case to make the transformation to the $\eps$-basis and back consistent. After solving the DE we arrive at 
\begin{align}
F_{4\mathrm{B}9}&= \frac{i \pi S_{\Gamma}^4}{\epsilon} \Bigg[\!\!-((\bar{z}-1) \bar{z})+\epsilon  \Bigg(\!-\frac{1}{3} \Big(36 (\bar{z}-1)+\pi ^2\Big) \bar{z}+2 (\bar{z}-1) H_ 0(\bar{z}) \bar{z}+2 H_ 2(\bar{z}) \bar{z}\nonumber \\
&+\Big(-3 \bar{z}^2+2 \bar{z}+1\Big) H_ 1(\bar{z}) \Bigg)+\epsilon ^2 \Bigg(\! -11 H_{1,1}(\bar{z}) \bar{z}^2+\frac{5 \pi ^2 \bar{z}^2}{6}+\frac{2}{3} \Big(36 (\bar{z}-1)+\pi ^2\Big) H_ 0(\bar{z}) \bar{z}\nonumber \\
&+(4 \bar{z}+11) H_ 2(\bar{z}) \bar{z}-6 H_ 3(\bar{z}) \bar{z}-4 (\bar{z}-1) H_{0,0}(\bar{z}) \bar{z}+4 H_{1,1}(\bar{z}) \bar{z}-4 H_{2,0}(\bar{z}) \bar{z}+14 H_{2,1}(\bar{z}) \bar{z}\nonumber \\
&-16 \zeta (3) \bar{z}-\frac{10 \pi ^2 \bar{z}}{3}+88 \bar{z}(1-\bar{z})+\Big(-36 \bar{z}^2+23 \bar{z}+13\Big) H_ 1(\bar{z})+\Big(6 \bar{z}^2-4 \bar{z}-2\Big) H_{1,0}(\bar{z})\nonumber \\
&+7 H_{1,1}(\bar{z})\Bigg)+ \qfto(\epstrip)\Bigg] \, .
\end{align}

\subsection*{Case 2: Asymptotic behaviour with Mellin-Barnes representation}
For the second case of integrals, calculating the asymptotic behaviour is not as straightforward as before: For more than two propagators, we generally need to introduce more than one Feynman parameter, leading to more complicated structures in the denominator, which can often not be expressed in terms of hypergeometric functions. To give an example for this type of integral, we now want to look at ${F}_{4\mathrm{B}29}$. After introducing Feynman parameters and using the phase-space parametrisation, we end up with the following expression
\begin{align}
F_{4\mathrm{B}29} &= 2\pi i S_{\Gamma}^4 \, \frac{2 \, \Gamma^2(1-\epsilon)\Gamma(1-2\epsilon) \Gamma(2-\epsilon) \Gamma(2+\epsilon)}{\Gamma(3-2\epsilon)\Gamma^2(1/2-\epsilon)} \int_0^1	du \ dv \ dx \ dw \ (z\bar{z})^{1-2\epsilon}  \nonumber \\[0.4em]
&\phantom{=} \times \int_0^1 dx_1 \ dx_2 \ dx_3 \ \frac{(u\bar{u})^{-1/2-\epsilon} v^{1-2\epsilon} (\bar{v}x\bar{x}w\bar{w})^{-\epsilon} \ \bar{x}_3^{-1-\epsilon} x_3}{[x_1 x_3 \bar{v} + v x_1 x_3 z \bar{w} + \bar{x}_3 + v x_1 x_2 x_3 \bar{z} + \bar{x}_1 x_2 x_3 \bar{z}]^{2+\epsilon}} (v \bar{x} \bar{z})^{-1}  \,.
\end{align}
As there is no analytical solution, we use several Mellin-Barnes transformations, allowing us to carry out the phase-space and Feynman integrations,
\begin{align}
F_{4\mathrm{B}29} =& \, 2\pi i S_{\Gamma}^4 \, \frac{2 \, \Gamma^4(1-\epsilon ) \Gamma(2-\epsilon )\Gamma (-\epsilon )}{\Gamma(1-2 \epsilon) \Gamma(3-2 \epsilon) \Gamma(-2 \epsilon )}  \, \int \frac{dz_1}{2 \pi i} \frac{dz_2}{2 \pi i} \frac{dz_3}{2 \pi i} \frac{dz_4}{2 \pi i} \ z^{z_2-2 \epsilon +1} \bar{z}^{z_3+z_4-2 \epsilon } \nonumber \\
& \times \Gamma (-z_1) \Gamma (-z_2) \Gamma (-z_3) \Gamma(-z_4) \Gamma (z_4+1)  \Gamma (z_1-\epsilon +1) \nonumber \\
& \times \Gamma(z_2-\epsilon +1) \Gamma (z_3+z_4+1) \Gamma (z_1+z_2+z_3+1)\Gamma (z_2+z_3-2\epsilon +1) \nonumber \\
& \times \frac{ \Gamma (-z_1-z_2-z_3-z_4-2 \epsilon -2) \Gamma (z_1+z_2+z_3+z_4+\epsilon +2)}{ \Gamma (z_2-2 \epsilon +2) \Gamma (z_3+z_4+2) \Gamma (z_1+z_2+z_3-3 \epsilon +2)} \,,
\end{align}
leaving us with only Mellin-Barnes integrations, and we will make use of the expansion properties of these integrals. For this we are using the programs \texttt{MB.m} and \texttt{MBasymptotics.m}~\cite{Czakon:2005rk}, to first expand the expression as a series in $\epsilon$ and then determine the asymptotic behaviour as $z\rightarrow 0,1$. After using the aforementioned tools, the four-fold representation simplifies greatly, leaving us with a maximum of two-fold representations to be calculated. The following Mellin-Barnes summations are either done with internal routines of \texttt{Mathematica} or with the additional package \texttt{HarmonicSums}~\cite{Ablinger:2009ovq}. After completing all the above steps, we arrive at the boundary condition
\begin{align}
F_{4\mathrm{B}29}|_{z\rightarrow 1} = \frac{i \pi S_{\Gamma}^4}{\epstrip \bar{z}} &\Bigg[ -\frac{\pi ^2}{6} +\eps  \left(\frac{2}{3} \pi ^2 \log (\bar{z})-7 \zeta (3)-\frac{\pi ^2}{3}\right) + \epssq \Big( -\frac{4}{3} \pi ^2 \log ^2(\bar{z}) \nonumber \\[0.5em]
&+\Big(28 \zeta (3) +\frac{4}{3} \pi ^2 \Big)\log (\bar{z})-14 \zeta(3)-\frac{5 \pi ^4}{18}-\frac{2\pi^2}{3}\Big) + \qfto(\epstrip)\Bigg] \, ,
\end{align}
which leads to the full solution
\begin{align}
F_{4\mathrm{B}29} = \frac{i \pi S_{\Gamma}^4}{\epstrip \bar{z}} &\Bigg[-\frac{\pi ^2}{6}+H_2(\bar{z})+\epsilon  \Bigg(\frac{2}{3} \pi ^2 H_ 0(\bar{z})+2 H_ 2(\bar{z})-4 H_3(\bar{z})-2 H_{2,0}(\bar{z})+7 H_{2,1}(\bar{z})\nonumber \\
&-7 \zeta (3)-\frac{\pi ^2}{3} \Bigg)+\epsilon ^2 \Bigg(\frac{4}{3} \Big(\pi ^2+21 \zeta (3)\Big) H_ 0(\bar{z})+\Big(4-2 \pi ^2\Big) H_ 2(\bar{z})-8 H_ 3(\bar{z})\nonumber \\
&+12 H_ 4(\bar{z})-\frac{8}{3} \pi ^2 H_{0,0}(\bar{z})-4 H_{2,0}(\bar{z})+14 H_{2,1}(\bar{z})-6 H_{2,2}(\bar{z})+4 H_{3,0}(\bar{z})\nonumber \\
&-28 H_{3,1}(\bar{z})-22 H_{2,1,0}(\bar{z})+39 H_{2,1,1}(\bar{z})-14 \zeta (3)-\frac{5 \pi ^4}{18}-\frac{2 \pi ^2}{3}\Bigg)+ \qfto(\epstrip)\Bigg]\,.
\end{align}

\subsection*{Case 3: Boundary condition from integration over {\boldmath $z$}}\label{sect:5Bconsts}

\begin{figure}[t]
\includegraphics[width=0.95\linewidth]{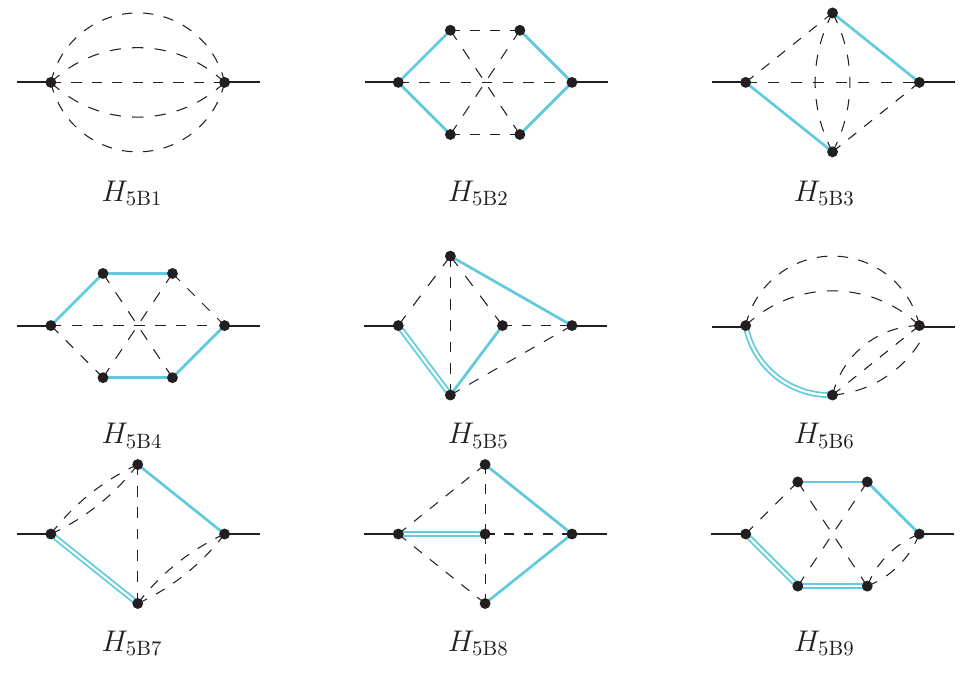}
\caption{The full set of master integrals we encounter in the reduction without the cut on the photon energy. The dashed lines indicate the cut propagators from reversed unitarity relations, the solid light blue (double-)lines indicate (massive) propagators.\label{fig:5BNoCutMIs}}
\end{figure}

This method is based on the idea that if we integrate a kernel over the full phase space without the cut, the result has to be equal to the kernel where we introduce a cut and integrate over the cut afterwards. This method was introduced in~\citere{Gituliar:2015iyq}  and has been successfully applied, for example, in~\citere{Moch:2021ult}. In our calculation we apply this method predominantly to the five-body master integrals to simplify the integration over the complicated five-body phase space. We introduce the notation $\widetilde{F}(\eps)$ for the cut-less integral, and thus write
\begin{equation}\label{eq:intBC}
\int_0^1 \, dz \, F_i(z,\epsilon) = \int_0^1 \, dz \, (\hat{T}(z,\epsilon) \vec{G}(z,\epsilon))_{i} = \widetilde{F}_i(\epsilon) \,,
\end{equation}
where the matrix $\hat{T}$ is the transformation matrix between the original and the $\epsilon$-basis while the vector $\vec{G}$ contains the solutions of the differential equations with undetermined constant parts. The structures $F_i(z,\epsilon)$ and $\widetilde{F}_i(\epsilon)$ are related to each other through the propagator originating from reverse unitarity described in section~\ref{sec:phasespacegeneral}. We can perform another IBP reduction to the cut-less integrals $\widetilde{F}_i(\eps)$ in eq.~\eqref{eq:intBC},
\begin{equation}\label{eq:intBCH}
\widetilde{F}_i(\epsilon) \mathrel{\stackrel{\makebox[0pt]{\mbox{\normalfont\tiny IBP}}}{=}}  \sum_j c_{ij}(\epsilon) \, H_j(\epsilon) \,,
\end{equation}
which yields the master integrals shown in figure~\ref{fig:5BNoCutMIs}. The massless integrals in figure~\ref{fig:5BNoCutMIs} were already computed in~\citere{Gituliar:2018bcr}. The $H_j(\eps)$ which contain massive propagators have not been known before and we computed them by means of their Mellin-Barnes representations. Finally, we compare~(\ref{eq:intBCH}) to the left-hand-side of eq.~\eqref{eq:intBC} to extract the boundary conditions. Our approach is illustrated by the following example,
\begin{align}\label{eq:5BBCprocedure}
\int_0^1 \, dz \,& \raisebox{-24.5pt}{\includegraphics[scale=0.4]{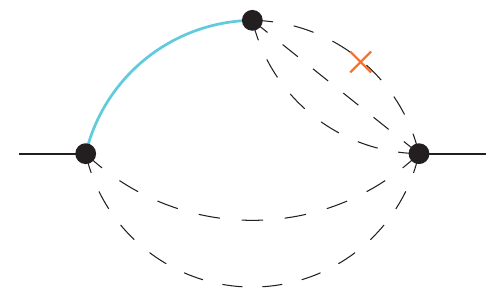}} = \raisebox{-24pt}{\includegraphics[scale=0.4]{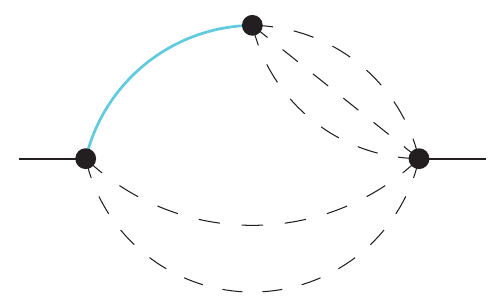}} \mathrel{\stackrel{\makebox[0pt]{\mbox{\normalfont\tiny IBP}}}{=}} g(\eps) \raisebox{-25pt}{\includegraphics[scale=0.4]{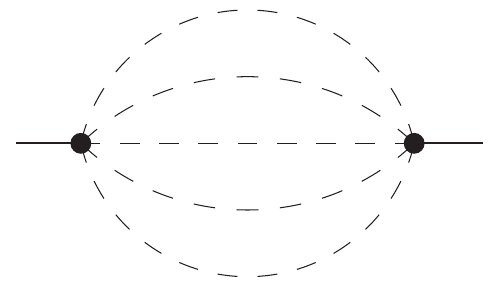}}\,,
\end{align}
where $g(\eps)$ is given by the IBP reduction. The results for all the master integrals are collected and can be found in electronic form in~\citere{url:GitHubRepo} and in the supplementary material at DOI.

\section{UV renormalisation and IR subtraction}
\label{sec:renormalisation}
After the computation of the bare contributions we have to renormalise the fields, the mass, and the operators to obtain UV-finite results. As the way on how to do the renormalisation is a priori not fixed, we can choose any scheme for renomalisation and we employ the on-shell-scheme for the mass and the fields. This has the benefit that all the light-quark $Z$-factors do not receive $\mathcal{O}(\alpha_s)$ corrections and we therefore only have to renormalise the $b$-quark. We can also neglect the photon wave function as the correction starts at $\mathcal{O}(\alpha_e \alpha_s)$ which lies beyond the scope of the current work. Therefore, the only $Z$-factors which are needed at $\mathcal{O}(\alpha_s)$ are $Z_m$ and $Z_h$ and $Z_1$ (of the $q\bar q\gamma$ vertex), which turn out to be equal at one loop in the on-shell scheme, 
\begin{equation}
    \label{eq:Zfactorpsi}
    Z_m = Z_1 = Z_h = 1 - \left(\frac{\alpha_s}{4\pi}\right) \frac{(3-2\eps)}{(1-2\eps)}\Gamma(\eps)e^{\eps(L_{\mu}+\gamma_E)}C_F \equiv 1 + \left(\frac{\alpha_s}{4\pi}\right) Z_h^{(1)},
\end{equation}
where $L_{\mu}=\log(\mu^2/m_b^2)$.

The renormalisation of the heavy quark spinor and the photon vertex are straightforward as the renormalisation is a simple product between the corresponding $Z$-factor and the tree-level amplitude. Diagrammatically this can be exemplary sketched as
\begin{equation}
\raisebox{-15pt}{\includegraphics[scale=0.38]{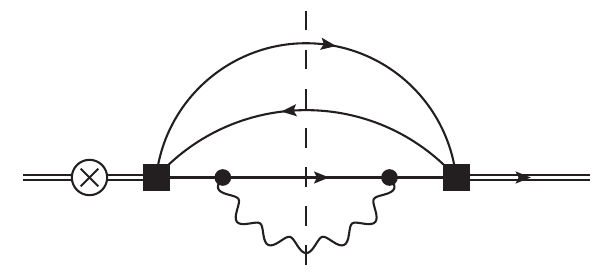}}= \left(\frac{\alpha_s}{4\pi}\right)\left[ \frac{1}{2} Z_{h}^{(1)}\raisebox{-15pt}{\includegraphics[scale=0.38]{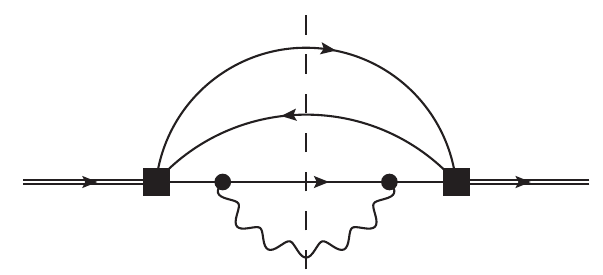}}\right]\,,
\label{eq:RenWF}
\end{equation}

\begin{equation}
\raisebox{-15pt}{\includegraphics[scale=0.38]{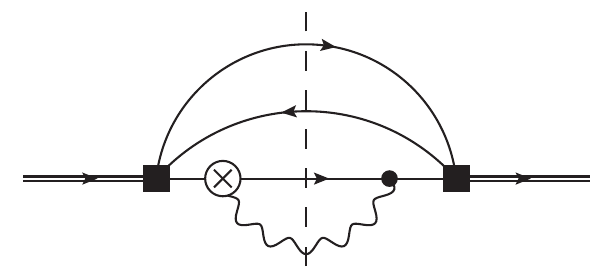}}=\left(\frac{\alpha_s}{4\pi}\right)\left[ Z_h^{(1)}\raisebox{-15pt}{\includegraphics[scale=0.38]{figs/Ren4B_Tree.pdf}}\right]\,,
\label{eq:RenVer}
\end{equation}
for the quark wave function and quark-antiquark-photon vertex, respectively. Finally, we have to renormalise the heavy quark propagator. The insertion of the counterterm changes the propagator expression. At $\mathcal{O}(\alpha_s)$ we observe that we obtain two different structures from this change. The first structure is again proportional to the tree-level amplitude while the second structure leads to a new type of diagram, with a dot on the propagator. Taking one of the occurring diagrams as an example, we can write this as

\begin{equation}
\raisebox{-15pt}{\includegraphics[scale=0.28]{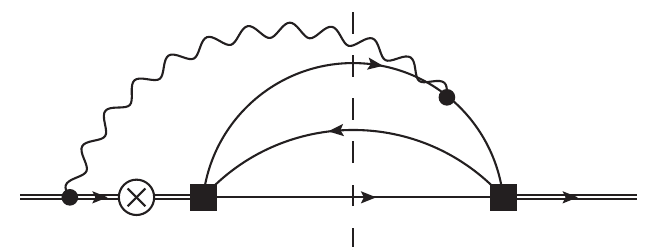}}=\left(\frac{\alpha_s}{4\pi}\right)\left[- Z_h^{(1)}\raisebox{-15pt}{\includegraphics[scale=0.28]{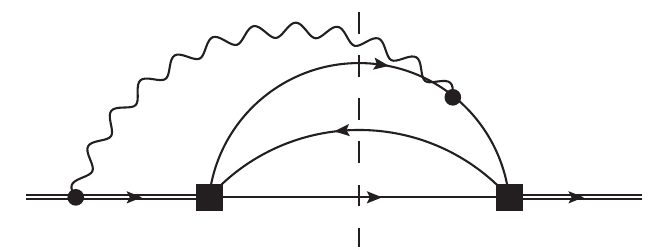}}-i \,Z_h^{(1)} \, m_b\raisebox{-15pt}{\includegraphics[scale=0.28]{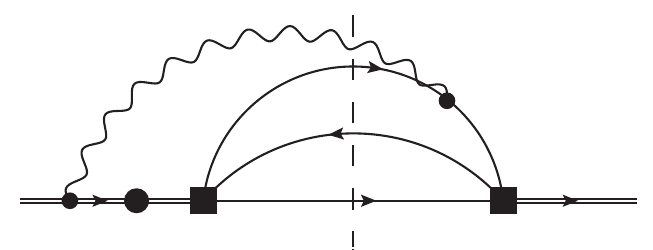}}\right]\,.
\label{eq:RenMass}
\end{equation}

Note that the structure proportional to tree-level amplitude has the same prefactor as the photon vertex, just with an opposite sign, and cancel once all diagrams are summed up. Thus we only have to deal with the wave-function counterterms and the second structure from the massive quark propagator, which we label $M_{ji}$ according to our master formula in eq.~\eqref{eq:Masterformula}. In the previous discussion it is always assumed that the counterterm is inserted to the left of the cut, however if it is inserted on the right-hand side the result is simply its complex conjugate.

Another source of UV divergences which have to be renormalised are the effective operators in eq.~\eqref{eq:operators}. The physical operators only span a complete basis in four dimensions, therefore we also have to incorporate a set of evanescent operators that give finite contributions in dimensional regularisation. The ones that are relevant in this work are~\cite{Gorbahn:2004my}
\begingroup
\allowdisplaybreaks
\begin{align}\label{eq:defineevops}
E_1 &= (\bar{s}_L \gamma_{\mu_1}\gamma_{\mu_2}\gamma_{\mu_3} T^a u_L)(\bar{u}_L \gamma^{\mu_1}\gamma^{\mu_2}\gamma^{\mu_3} T^a b_L)-16 P_1^u \nonumber \,,\\ 
E_2 &= (\bar{s}_L \gamma_{\mu_1}\gamma_{\mu_2}\gamma_{\mu_3} u_L)(\bar{u}_L \gamma^{\mu_1}\gamma^{\mu_2}\gamma^{\mu_3} b_L)-16 P_2^u \nonumber \,,\\ 
E_3 &= (\bar{s}_L \gamma_{\mu_1}\gamma_{\mu_2}\gamma_{\mu_3}\gamma_{\mu_4}\gamma_{\mu_5} b_L)\sum_{q}(\bar{q} \gamma^{\mu_1}\gamma^{\mu_2}\gamma^{\mu_3}\gamma^{\mu_4}\gamma^{\mu_5} q)+64 P_3-20 P_5  \nonumber \,,\\ 
E_4 &=  (\bar{s}_L \gamma_{\mu_1}\gamma_{\mu_2}\gamma_{\mu_3}\gamma_{\mu_4}\gamma_{\mu_5}T^a b_L)\sum_{q}(\bar{q} \gamma^{\mu_1}\gamma^{\mu_2}\gamma^{\mu_3}\gamma^{\mu_4}\gamma^{\mu_5} T^a q)+64 P_4-20 P_6\,.
\end{align}
The mixing matrix that is relevant for our calculation can be determined without computing additional diagrams. The matrix up to $\mathcal{O}(\alpha_s)$ and including the evanescent operators can be deduced from ref.~\cite{Gambino:2003zm} and is given by

\begin{equation}
Z=
\left(
\begin{array}{cccccccccc}
 -\frac{6}{N_c} & \frac{3 \left(N_c^2-1\right)}{2 N_c^2} & 0 & 0 & 0 & 0 & \frac{N_c^2-4}{4 N_c} &
   \frac{N_c^2-1}{4 N_c^2} & 0 & 0 \\[0.6em]
 6 & 0 & 0 & 0 & 0 & 0 & 1 & 0 & 0 & 0 \\[0.6em]
 0 & 0 & 0 & -10 & 0 & 1 & 0 & 0 & 0 & 0 \\[0.6em]
 0 & 0 & -\frac{5 \left(N_c^2-1\right)}{2 N_c^2} & -\frac{2 \left(2 N_c^2-5\right)}{N_c} &
   \frac{N_c^2-1}{4 N_c^2} & \frac{N_c^2-4}{4 N_c} & 0 & 0 & 0 & 0 \\[0.6em]
 0 & 0 & 0 & -64 & 0 & 10 & 0 & 0 & 0 & 1 \\[0.6em]
 0 & 0 & -\frac{16 \left(N_c^2-1\right)}{N_c^2} & -\frac{16 \left(N_c^2-4\right)}{N_c} & \frac{5
   \left(N_c^2-1\right)}{2 N_c^2} & \frac{N_c^2-10}{N_c} & 0 & 0 & \frac{N_c^2-1}{4 N_c^2} &
   \frac{N_c^2-4}{4 N_c}
\end{array}
\right)\,. \label{eq:Zij}
\end{equation}

The rows label the operators $\displaystyle\{ P_1^u,P_2^u,P_{3,\ldots,6},E_{1,\ldots,4}\}$, whereas the columns represent \\$\displaystyle\{ P_1^u,P_2^u,P_{3,\ldots,6}\}$ in that order. Note that \eqn{eq:Zij} does not contain the full one-loop renormalisation matrix from the effective weak Hamiltonian since part of its contribution was already used to renormalise the contributions in~\citere{Huber:2014nna}. The full one-loop renormalisation matrix is obtained by adding $\displaystyle \left\{-\frac{1}{3 N_c},\frac{2}{3},\frac{4}{3},\frac{2 f}{3}-\frac{2}{3 N_c},\frac{64}{3},\frac{20 f}{3}-\frac{32}{3 N_c}\right\}$  to the 4th column of~\eqref{eq:Zij}, see eq.~(2.48) of~\cite{Huber:2014nna}.

As before, we can express the procedure diagrammatically, for the insertions of the operator $P_i$ to the left of the cut, the diagrams we have to calculate take the following form:

\begin{align}
\raisebox{-25pt}{\includegraphics[scale=0.45]{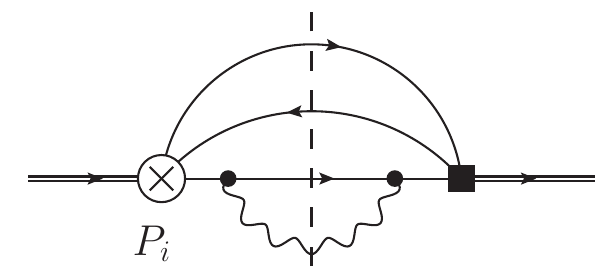}} = \left(\frac{\alpha_s}{4\pi \epsilon}\right) &\Bigg[ \sum_{j=1u,2u,3,4,5,6}   (Z_{PP})_{ij} \, \raisebox{-25pt}{\includegraphics[scale=0.45]{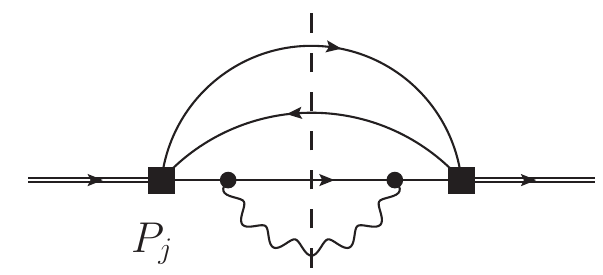}} \nonumber\\
& +  \sum _{j=1,2,3,4}  \, (Z_{PE})_{ij} \, \raisebox{-25pt}{\includegraphics[scale=0.45]{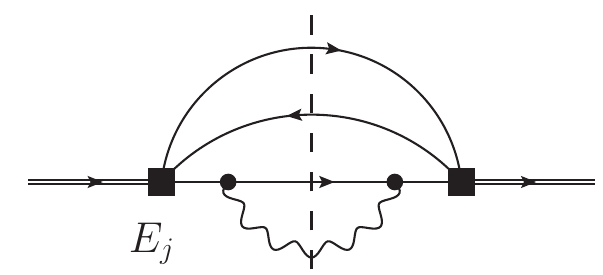}} \Bigg] \,,
\label{eq:RenOp}
\end{align}
where $Z_{PP}$ denotes the sub-matrix of $Z$ with the first 6 columns and $Z_{PE}$ the sub-matrix of the last 4 columns. The operators denoted by $E_j$ are the evanescent operators defined in eq.~\eqref{eq:defineevops}. We point out that the diagrams with evanescent operator insertion only affect the finite part of the result as the diagrams themselves vanish as $\ep \rightarrow 0$.

After performing the aforementioned steps our calculation is now UV-finite. However, we still have to deal with remaining IR-divergences. These singularities arise from the regions of phase
space where the radiated photon is collinear to the light quarks and are an artifact of the latter being treated massless. Treating them as massive would avoid these collinear divergences at the price of making the integrals more involved. In order to deal with the collinear divergences we employ the same method as in ref.~\cite{Huber:2005ig,Huber:2014nna,Kaminski:2012eb}, where the differential decay width reads
\begin{equation}
    \label{eq:DecayShiftprescription}
    \frac{\mathrm{d}\Gamma_{m}}{\mathrm{d}z}=\frac{\mathrm{d}\Gamma_{\eps}}{\mathrm{d}z}+\frac{\mathrm{d}\Gamma_{\text{shift}}}{\mathrm{d}z}\,,
\end{equation}
where $1-z\!=2E_{\gamma}/m_b$. The first term on the RHS above is the dimensionally regulated expression, while the second one converts the dimensional regulators to logarithms of the light quark mass $m_q$. The contribution $\mathrm{d}\Gamma_{\text{shift}}/\mathrm{d}z$ is related to the difference between emitting a collinear photon from a massive and massless quark.

At LO the IR-subtraction kernel, denoted as $S_0$ in eq.~\eqref{eq:Masterformula},  is given by the difference between the integrated massive~\cite{Catani:2000ef} and massless~\cite{Altarelli:1977zs} quark to photon splitting function at $\mathcal{O}(\alpha_e)$ and reads
\begin{equation}
\label{eq:LOSubKernel}
    S_0=\frac{\alpha_e}{4\pi}\frac{2\pi e^{\gamma_E \eps}(1-\eps)(1+\bar{z}_{\gamma}^2)\csc(\pi  \eps)}{z_{\gamma}\Gamma(1-\eps)}\left(\frac{m_q z_{\gamma}}{\mu}\right)^{-2\eps}\,,
\end{equation}
where $z_{\gamma}$ is the momentum fraction of the photon and $m_q$ is the light quark mass. We then express the momentum fraction $z_{\gamma}$ in terms of the cut on the photon energy $z$ through the relation
\begin{equation}
\label{eq:CutPhotonRel}
    z_{\gamma}=\frac{\bar{z}}{1-s_{kl}}\, ,
\end{equation}
where $k$ and $l$ are the momentum labels of the two final-state quarks that are not involved in the collinear splitting. Thus we are able to write the final shift-relation for the tree-level contribution as
\begin{align}
    \label{eq:TreeLevelShift}
    \left(\frac{\mathrm{d}\Gamma^{T,0}_{\text{shift}}}{\mathrm{d}z}\right)_{ij}&=T^{\slashed{\gamma_c}}_{ij} \otimes S_{0}=\frac{1}{2m_b}\frac{1}{2N_c}\int dPS_3 \,\,\mathcal{T}_{ij}(s_{kl})\frac{\alpha_e}{2\pi \bar{z}}\bigg{\{}Q_1^2\left[1+\frac{(z-s_{23})^2}{(1-s_{23})^2}\right]\nonumber\\
    &\times\left[\frac{\pi e^{\gamma_E \eps} (1-\eps)\csc(\pi \eps)}{\Gamma(1-\eps)}\left(\frac{m_{q_1}(1-z)}{\mu(1-s_{23})}\right)^{-2\eps}\right]\Theta(z-s_{23})+\text{cyclic}\bigg{\}}\,,
\end{align}
where $i$ and $j$ denote again the operator insertions and we label the photon momentum by $p_4$ and the momentum of quark $q_k$ by $p_k$. In eq.~\eqref{eq:TreeLevelShift}, cyclic means that we have to sum over the IR-subtraction kernels being attached to every of the three cut quark lines with the respective charges, masses and momenta adjusted. The terms $\mathcal{T}_{ij}(s_{kl})$ denotes the spin-summed squared matrix element of the $b$ quark decaying into the three light quarks at tree-level. Note that the integration is now over the entire phase-space volume and the $z$-dependence is introduced at the level of the integrand.

In order to properly subtract the IR singularities at NLO we now have to consider the LO IR-subtraction kernel convoluted with the NLO matrix elements, both with a virtual gluon and with an on-shell gluon. In addition we also have to consider the NLO IR-subtraction kernel convoluted with the tree-level matrix elements. The first contribution are the diagrams with an additional virtual gluon as these are nearly identical to the tree-level contribution. The shift-relation is equivalent to eq.~\eqref{eq:TreeLevelShift} once the matrix element $\mathcal{T}_{ij}(s_{kl})$ is replaced with $\mathcal{V}_{ij}(s_{kl})$ which denotes the associated one-loop contribution. The calculation then follows the exact same steps as for the tree-level contribution. In the case of an additional on-shell gluon emission we have to adjust the relation between the photon momentum fraction and the cut on the photon energy. In this case the relation now reads
\begin{equation}
\label{eq:CutPhotonRel5Body}
    z_{\gamma}=\frac{\bar{z}}{1-s_{kl5}}\, ,
\end{equation}
where $k$ and $l$ are again the momentum labels of the two final-state quarks that are not involved in the collinear splitting and the label "$5$" represents the gluon. By employing the adjusted relation and labelling the gluon momentum by $p_5$, we can now write the shift-relation for the additional gluon emission as
\begin{align}
    \label{eq:NLORealEmiShift}
    \left(\frac{\mathrm{d}\Gamma^{R,0}_{\text{shift}}}{\mathrm{d}z}\right)_{ij}&=R^{\slashed{\gamma_c}}_{ij} \otimes S_{0}=\frac{1}{2m_b}\frac{1}{2N_c}\int dPS_4 \,\,\mathcal{R}_{ij}(s_{kl})\frac{\alpha_e}{2\pi \bar{z}}\bigg{\{}Q_1^2\left[1+\frac{(z-s_{235})^2}{(1-s_{235})^2}\right]\nonumber\\
    &\times\left[\frac{\pi e^{\gamma_E \eps} (1-\eps)\csc(\pi \eps)}{\Gamma(1-\eps)}\left(\frac{m_{q_1}(1-z)}{\mu(1-s_{235})}\right)^{-2\eps}\right]\Theta(z-s_{235})+\text{cyclic}\bigg{\}}\,,
\end{align}
where $\mathcal{R}_{ij}(s_{kl})$ is the spin-summed squared matrix element of the $b$ quark decaying into a gluon and three light quarks at tree-level. The rest of the notation is equivalent to the one introduced in eq.~\eqref{eq:TreeLevelShift}. As the $z$-dependence is again introduced at the level of the integrand, the integration over the entire volume of the four-particle phase space leaves us more freedom in choosing the order in which the integrations are carried out.

With the introduction of $\mathcal{V}_{ij}(s_{kl})$ we have reintroduced UV-divergences in our calculation which again have to be renormalised. The procedure is identically to the one for the full amplitudes. However as there are no photons in these kinds of diagrams we do not have to deal with any vertex, propagator or mass renormalisation. The only parts that are relevant are the quark wave-functions and the operator insertions. As we have seen from eqs.~\eqref{eq:Zfactorpsi} and~\eqref{eq:RenOp} these counterterms are multiplicative and therefore no new diagrams have to be calculated.

The final piece which is required to cancel all the IR-singularities is the NLO IR-subtraction kernel. Requiring that the kernel subtracts all the IR poles, we only need the pole structure of the subtraction kernel since the matrix element with which it is convoluted is finite. The NLO IR-subtraction kernel is given by
\begin{align}
\label{eq:NLOIRSubtraction}
S_1=\frac{\alpha_e\alpha_s}{(4\pi)^2}\, C_F  \, \left(\frac{m_q}{\mu}\right)^{-4\eps} \left[\frac{1}{2\eps^2} \, \left(P^{(0)}_{q\to q} \otimes P^{(0)}_{q\to \gamma}\right)(z_{\gamma}) - \frac{1}{2\eps} R^{(1)}_{q\to \gamma}(z_{\gamma})\right]\, ,
\end{align}
where $P^{(0)}_{q\to i}$ are the coefficients of the DGLAP splitting functions which are given by
\begin{alignat}{4}\label{eq:DGLAPSF}
P^{(0)}_{q\to q}(z_{\gamma}) &= \, \frac{1+z_{\gamma}^2}{(1-z_{\gamma})_+} + \frac{3}{2} \, \delta(1-z_{\gamma}) \, , \,\qquad&&  
P^{(0)}_{q\to \gamma}(z_{\gamma}) &&= \, \frac{1+(1-z_{\gamma})^2}{z_{\gamma}} \, .
\end{alignat}
We furthermore introduced a short-hand notation for the convolution
\begin{equation}
    \label{eq:Def_Conv}
    \left(P^{(0)}_{q\to q} \otimes P^{(0)}_{q\to \gamma}\right)(z_{\gamma})=\int_{z_{\gamma}}^1 \frac{\mathrm{d}\xi}{\xi}\,P^{(0)}_{q\to q}(\xi)P^{(0)}_{q\to \gamma}\left(\frac{z_{\gamma}}{\xi}\right)\,,
\end{equation}
and defined the plus-distribution via
\begin{equation}
    \label{eq:Def_PD}
    \int\limits_0^1 \! dz \, \frac{f(z)}{(1-z)_+}  = \int\limits_0^1 \! dz \, \frac{f(z)-f(1)}{1-z} \, .
\end{equation}
While the leading pole follows the DGLAP equation, the subleading contribution does not and instead is given by 
\begin{align}
R^{(1)}_{q\to \gamma}(z_{\gamma}) =& \, \frac{1+(1-z_{\gamma})^2}{z_{\gamma}} \left(16 \ln (1-z_{\gamma}) \ln(z_{\gamma}) -4 \, \ln^2(1-z_{\gamma}) + \frac{16}{3} \, \pi ^2 -30\right)\nonumber \\[0.5em]
&-6 (z_{\gamma}-2) \ln^2(z_{\gamma})+\frac{8 (z_{\gamma}-2)^2}{z_{\gamma}} \,  \ln(1-z_{\gamma})-2 (z_{\gamma}-16) \, \ln(z_{\gamma}) +\frac{2 \left(3z_{\gamma}+10\right)}{z_{\gamma}}\, .
\end{align}
Similar to the tree-level contribution, we can use the expression from eq.~\eqref{eq:CutPhotonRel} and obtain for the shift-relation
\begin{align}
    \label{eq:NLOSubShift}
    \left(\frac{\mathrm{d}\Gamma^{T,1}_{\text{shift}}}{\mathrm{d}z}\right)_{ij}&=T^{\slashed{\gamma_c}}_{ij} \otimes S_{1}=\frac{1}{2m_b}\frac{1}{2N_c}\int dPS_3 \,\,\mathcal{T}_{ij}(s_{kl})\frac{\alpha_e \alpha_s}{(4\pi)^2}C_F\bigg{\{}Q_1^2\left(\frac{m_{q_1}}{\mu}\right)^{-4\eps}\nonumber\\
    &\times\left[\frac{1}{2\eps^2} \, \left(P^{(0)}_{q\to q} \otimes P^{(0)}_{q\to \gamma}\right)\left(\frac{\bar{z}}{1-s_{23}}\right) - \frac{1}{2\eps} R^{(1)}_{q\to \gamma}\left(\frac{\bar{z}}{1-s_{23}}\right)\right]\Theta(z-s_{23})+\text{cyclic}\bigg{\}}\,,
\end{align}
where we again employed the same notation as in eq.~\eqref{eq:TreeLevelShift}.

%%%%%%%%%%%%%%%%%%%%%%%%%%%%%%%%%%%%%%%%%%%%%%%%%%%%%%%%%%%%%%%%%

\section{Analytic results}
\label{sec:analyticresults}

After all ingredients have been plugged into the master formula~\eqref{eq:Masterformula}, all poles in the dimensional regulator $\eps$ cancel analytically and we are left with a finite result, which subsequently gets integrated over $z \in [0,\ldots,\delta]$. The structure of our expression is such that we obtained all entries $G_{ij}^{(1)}(\mu,\delta)$ completely analytically in terms of harmonic and Goncharov polylogarithms. Moreover, it turns out to be efficient to organise the result in terms of building blocks $T_{n}$ which appear repeatedly in the $G_{ij}^{(1)}(\mu,\delta)$. We furthermore subdivided the results into $G_{ij}^{(I)}$ for the diagrams containing two and $G_{ij}^{(II)}$ for the diagrams containing only a single Dirac trace.

Our analytic results can be found on \texttt{GitHub}~\cite{url:GitHubRepo} and as supplementary material at DOI, where the entries of the resulting matrix can be constructed using \texttt{CombineGij.nb}, and the integrals we calculated in sec.~\ref{Sec:barecalculation} can be found in \texttt{Integrals.nb}. To illustrate the general form, we give an explicit example here. One of the simpler blocks we encounter is the part of $G_{13}^{(II)}$ that is proportional to the charge factor $Q_d^2$ and the colour factor $C_F C_A$,
\begin{equation}
    G^ {(II)}_{13,Q_d^2,C_F C_A}=-\frac{T_{15}}{4}L_{\mu}L_q+2T_7L_{\mu}+ \frac{\left(2T_{13}-75T_{15}\right)}{360}L_q-\frac{\left(T_{17}+336 T_7\right)}{240},
\end{equation}
containing the building blocks
\begin{align*}
    T_{7}&=\frac{1}{288} \big(61 \bar{\delta }^4-232 \bar{\delta }^3+330 \bar{\delta }^2-760 \bar{\delta }+601\big)+\frac{17}{12} H_0\big(\bar{\delta }\big)\\
	&+\frac{1}{12} \big(\bar{\delta }^4-4 \bar{\delta }^3+6 \bar{\delta }^2-10 \bar{\delta }+7\big) H_1\big(\bar{\delta }\big)+\frac{1}{2} H_2\big(\bar{\delta }\big)-\frac{\pi^2}{12} \,, \\[10pt]
T_{13}&= 5 \pi ^2 \big(3 \bar{\delta }^2+6 \bar{\delta }+2\big)-\frac{5}{4} \big(53 \bar{\delta }^4-288 \bar{\delta }^3+394 \bar{\delta }^2-792 \bar{\delta }+633\big)\\
	&-30\big(3 \bar{\delta }^2+6 \bar{\delta }+2\big) H_2\big(\bar{\delta }\big)+5 \big(6 \bar{\delta }^4-20 \bar{\delta }^3+27 \bar{\delta }^2-111\big)
   H_0\big(\bar{\delta }\big)\\
   &+5 \big(4 \bar{\delta }^3+27 \bar{\delta }^2-31\big) H_1\big(\bar{\delta }\big)\,, \\[10pt]  	
T_{15}&=\frac{2}{3} \big(\bar{\delta }^4-4 \bar{\delta }^3+6 \bar{\delta }^2-10 \bar{\delta }+7\big)+4 H_0\big(\bar{\delta }\big) \,, \\[10pt] 
T_{17}&= -\frac{20}{3} \big(6 \bar{\delta }^2-20 \bar{\delta }+27\big) \bar{\delta }^2 H_{0,0}\big(\bar{\delta }\big)-\frac{20}{3} \big(2 \bar{\delta }^3-3 \bar{\delta}^2+26 \bar{\delta }-28\big) H_{1,0}\big(\bar{\delta }\big)\\
   &+\frac{40}{3} \big(4 \bar{\delta }^3+27 \bar{\delta }^2-31\big) H_{1,1}\big(\bar{\delta }\big)+480 H_{2,0}\big(\bar{\delta }\big)-80 \big(3 \bar{\delta }^2+6 \bar{\delta }+2\big) H_{2,1}\big(\bar{\delta }\big)\\
   &+\frac{1}{36} \big(-12097 \bar{\delta }^4+65204\bar{\delta }^3-79020 \bar{\delta }^2+244300 \bar{\delta }-218867\big)\\
   &+\frac{4}{9} \pi ^2 \big(-30 \bar{\delta }^4+110 \bar{\delta }^3-90 \bar{\delta }^2+270 \bar{\delta }+363\big)+40 \zeta (3) \big(3 \bar{\delta }^2+12 \bar{\delta }+28\big)\\
   &+120 \bar{\delta }^2 H_3\big(\bar{\delta }\big)+\Big(\frac{1315 \bar{\delta }^5-5495 \bar{\delta }^4+9070 \bar{\delta }^3-10770 \bar{\delta }^2-25704 \bar{\delta }+31704}{9 \big(\bar{\delta }-1\big)}\\
   &+\frac{80}{3} \pi ^2\Big) H_0\big(\bar{\delta }\big)+\frac{1}{9} \big(-947 \bar{\delta }^4+6808 \bar{\delta }^3-4062 \bar{\delta }^2+14180 \bar{\delta }-15979\big) H_1\big(\bar{\delta }\big)\\
   &+\frac{4}{3} \big(30 \bar{\delta }^4-140 \bar{\delta }^3-45\bar{\delta }^2-190 \bar{\delta }-886\big) H_2\big(\bar{\delta }\big)\,.
\end{align*}
Here, logarithms containing the renormalisation scale $\mu$ are abbreviated by $ L_{\mu} = \log \left( {\mu^2}/{m_b^2} \right)$, and the collinear logarithms by $ L_{q} = \log \left({m_b^2}/{m_q^2} \right)$.\footnote{Note that this notation is different from the one used in~\cite{Huber:2014nna}.} For the entries that contain more than one Dirac trace, the results also contain the number $n_0$ of active light flavours which in our case is set to $n_0=3$ for up, down and strange.

%%%%%%%%%%%%%%%%%%%%%%%%%%%%%%%%%%%%%%%%%%%%%%%%%%%%%%%%%%%%%%%%%%
%%%%%%%%%%%%%%%%%%%%%%%%%%%%%%%%%%%%%%%%%%%%%%%%%%%%%%%%%%%%%%%%%%

\section{Numerical results}
\label{sec:numericalresults}

In order to investigate the size of the correction computed in the present work we remind the reader of eq.~\eqref{eq:PartonicDecayRate4B5B}, which we use to define the quantity $\Delta\Gamma$,
\begin{equation}
  \Delta \Gamma \equiv  \Gamma(b \rightarrow s q\bar{q}\gamma)_{E_{\gamma}>E_0}+\Gamma(b \rightarrow s q\bar{q}g\gamma)_{E_{\gamma}>E_0}= \Gamma_0 \sum_{i,j} \mathcal{C}_i^{\text{eff} \, *}(\mu) \, \mathcal{C}_j^{\text{eff}}(\mu) \,  \widehat{G}_{ij}(\mu,z_c,\delta) \,. \label{eq:bsqqgrate10x10}
\end{equation}
The sum runs over $i,j = 1u,2u,3,\ldots,6,1c,2c,7,8$, i.e.\ $\widehat{G}_{ij}$ is a hermitian $10 \times 10$ matrix. The Wilson coefficients are expanded in the strong coupling as follows,
\begin{align}
\C{i}^{\text{eff}}(\mu) & = \C{i}^{\text{(0)eff}}(\mu) + \frac{\alpha_s(\mu)}{4\pi} \C{i}^{\text{(1)eff}}(\mu) + \ord{\alpha_s^2} \, .
\end{align}
The one-loop expressions for the Wilson coefficients and anomalous dimension can be found in~\citere{Chetyrkin:1996vx}. To obtain their numerical values, we use the input values collected in table~\ref{tab:inputs}. The strong coupling is evolved to a scale $\mu$ using four-loop running as implemented in {\tt RunDec.m}~\cite{Herren:2017osy}, the mass of the top quark is converted to the \MSbar~scheme using four-loop accuracy, again using {\tt RunDec.m}. This yields for the central values of the relevant entries at $\mu \sim m_b/2 =2.5$~GeV
\eqa{
\C{i}^{(0)\text{eff}} &=&
 (0.804 \,\lambda_u , -1.060 \,\lambda_u , -0.0122 , -0.1204 , 0.0011 , 0.0026 ,   \nonumber \\[0.3em] && \; 0.804 \,\lambda_c , -1.060 \,\lambda_c , -0.367 , -0.170 ) \, , \quad \\
\C{i}^{(1)\text{eff}} &=&
(-15.454 \,\lambda_u,2.061 \,\lambda_u,0.101,-0.398,-0.0210,-0.0155, \nonumber \\[0.3em] && \; -15.454 \,\lambda_c,2.061 \,\lambda_c,-,-) \, .
}

%%%%%%%%%%%%%%%%%%%%%%%%%%%%%%%%
\begin{table}[t]
	\begin{center}
		\begin{displaymath}
		\begin{tabular}{|l|l|}
		\hline\spp
        $m_t^{\text{pole}}=(172.4\pm 0.7)$~GeV & $\alpha_s^{(5)}(M_Z)=0.1180\pm 0.0009$ \\ \spp
        $m_b^{\text{pole}}=(4.78\pm 0.06)$~GeV & $M_Z = 91.1880\;\gev$    \\ \spp
        $m_b^{\text{1S}}=(4.65\pm 0.03)$~GeV & $M_W = 80.3692\;\gev$  \\ \spp
        $\overline{m}_b(\overline{m}_b)=(4.183\pm 0.007)$~GeV &  $\mu_0 = (160 \pm 80)$~GeV  \\ \spp
        $m_c^{\text{pole}}=(1.67\pm 0.07)$~GeV & $\mu = 2.5^{+2.5}_{-0.5}$~GeV  \\ \spp
        $\overline{m}_c(\overline{m}_c)=(1.2730\pm 0.0046)$~GeV &  $z_c = m_c^2/m_b^2 = 0.12 \pm 0.03 $ \\ \spp
        $\lambda_u = -0.0086 + 0.0186 \, i$  & $E_0 = 1.6$~GeV  \\ \spp
        $\lambda_c = -0.9914 - 0.0186 \, i$  &   \\ 
        \hline
		\end{tabular}
		\end{displaymath}
	
		\vspace*{-10pt}
		
		\caption{Numerical input parameters used in the numerical analysis. Numbers are taken from PDG~\cite{ParticleDataGroup:2024cfk} and CKMfitter~\cite{Charles:2004jd,ValeSilva:2024jml}. Numbers without uncertainties stem from external parameters (like $E_0$) or have error bars that are negligible for our purposes. See text for further explanations.\label{tab:inputs}}
	\end{center}  
\end{table}
%%%%%%%%%%%%%%%%%%%%%%%%%%%%%%%%

The matrix $\widehat{G}$ is expanded in the same way as the Wilson coefficients (see eq.~\eqref{eq:defGij}),
\begin{equation}
 \widehat{G}_{ij}(\mu,z_c,\delta)= \widehat{G}_{ij}^{(0)}(\delta)+ \frac{\alpha_s(\mu)}{4 \pi} \widehat{G}_{ij}^{(1)}(\mu,z_c,\delta) + \mathcal{O}(\alpha_s^2) \,.
\end{equation}
The LO matrix $\widehat{G}_{ij}^{(0)}(\delta)$~\cite{Kaminski:2012eb} is independent of $\mu$ and $z_c=m_c^2/m_b^2$ and only its upper left $6\times6$ block is populated. The matrix $\displaystyle\widehat{G}_{ij}^{(1)}(\mu,z_c,\delta) \equiv {G}_{ij}^{(1)}(\mu,z_c,\delta)+{G}_{ij}^{(1)}(\mu,\delta)$ is comprised of the sum of the contribution  computed in~\citere{Huber:2014nna} and the one evaluated in the present work. While the former depends on $\mu$ and $z_c$ and populates all but the lower right $4\times 4$ block, the latter is again independent of $z_c$ and non-zero only in the upper left $6\times6$ block. With this, we get for the quantity $\Delta\Gamma/\Gamma_0$, expanded in $\alpha_s$,
\begin{eqnarray}
\frac{\Delta\Gamma}{\Gamma_0} = \sum_{
\begin{minipage}{17mm}
\scriptsize\flushright
$i,j=1u,2u,$\\[-0.8mm]$3,\ldots,6$
\end{minipage}
} \C{i}^{\text{(0)eff}\,*}\, \C{j}^{\text{(0)eff}} \, \widehat{G}_{ij}^{(0)} &+& \frac{\alpha_s(\mu)}{4\pi} \Bigg[ \sum_{
\begin{minipage}{17mm}
\scriptsize\flushright
$i,j=1u,2u,$\\[-0.8mm]$3,\ldots,6$
\end{minipage}
}
\left(\C{i}^{\text{(1)eff}\,*} \, \C{j}^{\text{(0)eff}}+\C{i}^{\text{(0)eff}\,*} \, \C{j}^{\text{(1)eff}}\right) \,\widehat{G}_{ij}^{(0)} \nonumber \\[2mm]
&& \qquad \quad + \sum_{
\begin{minipage}{17mm}
\scriptsize\flushright
$i,j=1u,2u,$\\[-0.8mm]  \hspace*{-8pt} $3,\ldots,6,1c,2c,7,8$
\end{minipage}
} \C{i}^{\text{(0)eff}\,*}\,\C{j}^{\text{(0)eff}} \,\widehat{G}_{ij}^{(1)}\Bigg] + \ord{\alpha_s^2}\, . 
\label{eq:deltagammaexpanded}
\end{eqnarray}
We checked explicitly that this expression is renormalisation-group (RG) invariant up to the order we are working, i.e.\ $\text{d}\Delta\Gamma/\text{d}\log\mu =  \ord{\alpha_s^2}$.

In order to estimate the numerical size of the total four- and five-body correction through to NLO in QCD, we first plug into $\Delta\Gamma/\Gamma_0$ the central values from table~\ref{tab:inputs} (corresponding to $\delta = 1-2E_0/m_b=0.3305$). To numerically evaluate the polylogarithms in our result, we use {\texttt{PolyLogTools}}~\cite{Duhr:2019tlz}. For the size of the collinear logarithm $\log(m_b/m_q)$, a value of $m_q \sim m_s \sim 100$~MeV leads to $m_b/m_q \sim 50$, whereas the choice of a constituent mass $m_q \sim 250$~MeV results in $m_b/m_q \sim 20$. Ratios of meson masses such as $m_B/m_\pi \sim 35$ or $m_B/m_K \sim 10$ lead to a similar range. Also fragmentation functions have been used in the past to quantify the uncertainty associated with collinear logarithms, see e.g.~\cite{Kapustin:1995fk,Ferroglia:2010xe,Asatrian:2013raa}. For the various ratios, we obtain
\begin{align}
\Delta\Gamma/\Gamma_{0}{}_{\big|_{m_q=m_b/50}} & = \, 0.0256~\% \, , & \Delta\Gamma/\Gamma_{0}{}_{\big|_{m_q=m_b/35}} &  = \, 0.0194~\%\, , \\[0.5em]
\Delta\Gamma/\Gamma_{0}{}_{\big|_{m_q=m_b/20}} & = \, 0.0058~\%\, , & \Delta\Gamma/\Gamma_{0}{}_{\big|_{m_q=m_b/10}} & = \, -0.0175~\% \, .
%
%\end{align}
\intertext{The total size of the correction is in the per mille range or slightly below. However, relative to the leading-order two-body $b\to s \gamma$ decay width, $\Gamma_0 |\C{7}^{\text{(0)eff}}|^2$, it is in the range of a percent or slightly below since $|\C{7}^{\text{(0)eff}}|^2 \sim 0.1$,}
%\begin{align}
\Delta\Gamma/\Gamma_{0}|\C{7}^{\text{(0)eff}}|^2{}_{\big|_{m_q=m_b/50}} & = \, 0.1899~\% \, , & \Delta\Gamma/\Gamma_{0}|\C{7}^{\text{(0)eff}}|^2{}_{\big|_{m_q=m_b/35}} & = \, 0.1436~\% \, , \\[0.5em]
\Delta\Gamma/\Gamma_{0}|\C{7}^{\text{(0)eff}}|^2{}_{\big|_{m_q=m_b/20}} & = \, 0.0429~\% \, , & \Delta\Gamma/\Gamma_{0}|\C{7}^{\text{(0)eff}}|^2{}_{\big|_{m_q=m_b/10}} & = \, -0.1297~\% \, .
\end{align}
Both, the contribution calculated in~\citere{Huber:2014nna} (called NLO4B in the following) and the present one (NLO4B+5B) are negative and thus lower the value found at LO. The size of the NLO4B+5B correction is expected to be smaller compared to NLO4B due to CKM suppression, small Wilson coefficients of QCD penguin operators, and phase-space suppression. However, this suppression is partially lifted by the collinear logarithm $\log(m_b^2/m_q^2)$ which appears linearly in NLO4B but which shows up quadratically in the NLO4B+5B expressions. To illustrate this numerically, we give the various contributions to $\Delta\Gamma/\Gamma_0$ individually for $m_q=m_b/50$ and $m_q=m_b/20$ at $\mu=2.5$~GeV (all numbers in percent),
\begin{align}
\Delta\Gamma/\Gamma_{0}{}_{\big|_{m_q=m_b/50}}[\%] & = (0.0600)_{\text{LO}} - (0.0168)_{\text{NLO4B}} - (0.0179)_{\text{NLO4B+5B}} \, , \\[0.6em]
\Delta\Gamma/\Gamma_{0}{}_{\big|_{m_q=m_b/20}}[\%] & = (0.0274)_{\text{LO}} - (0.0158)_{\text{NLO4B}} - (0.0061)_{\text{NLO4B+5B}} \, .
\end{align}
%%%%%%%%%%%%%%%%%%%%%%%%%%%%%%%%
\begin{figure}[t]
    \centering
    \includegraphics[width=0.483\textwidth]{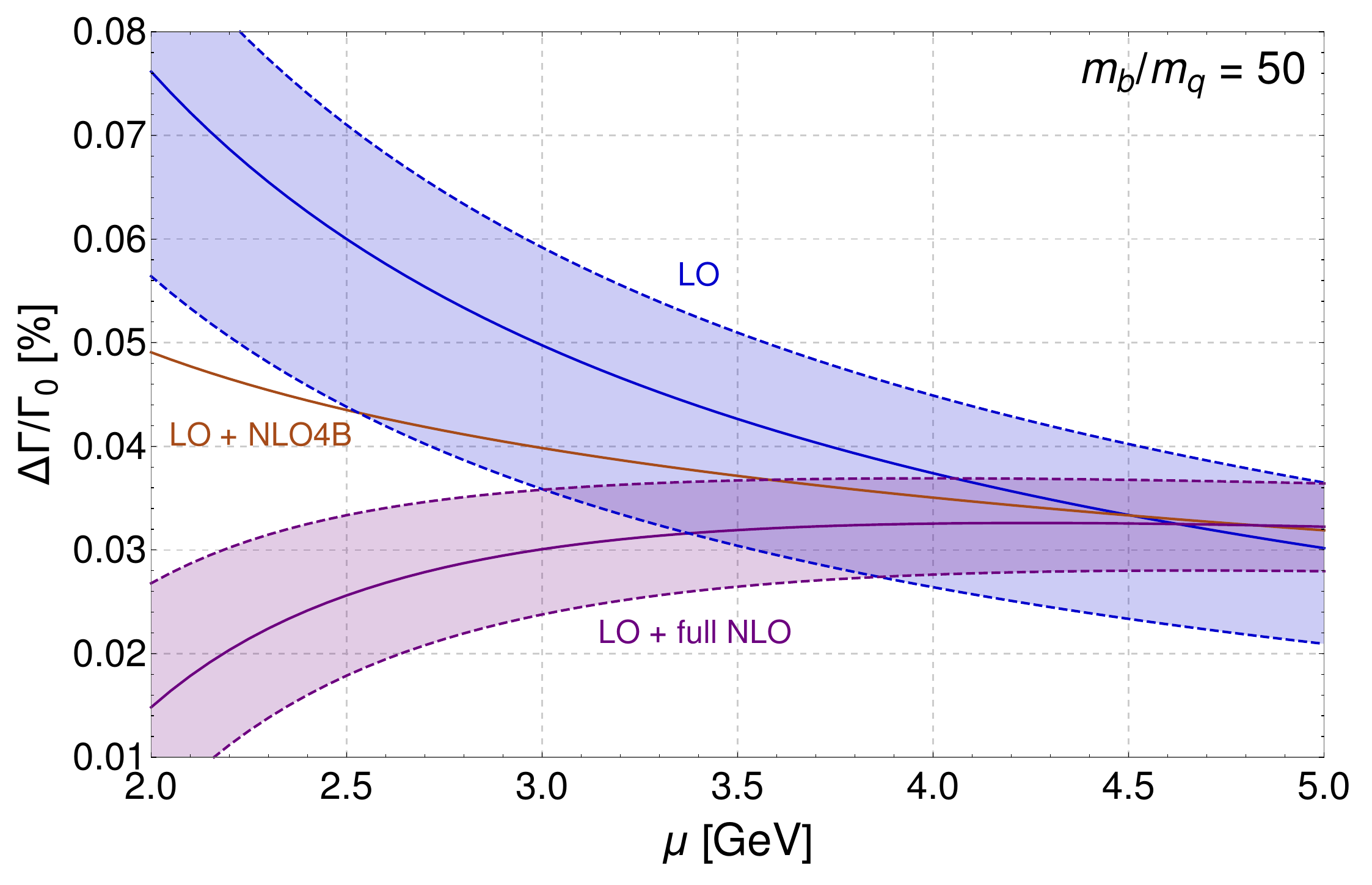}
    \includegraphics[width=0.49\textwidth]{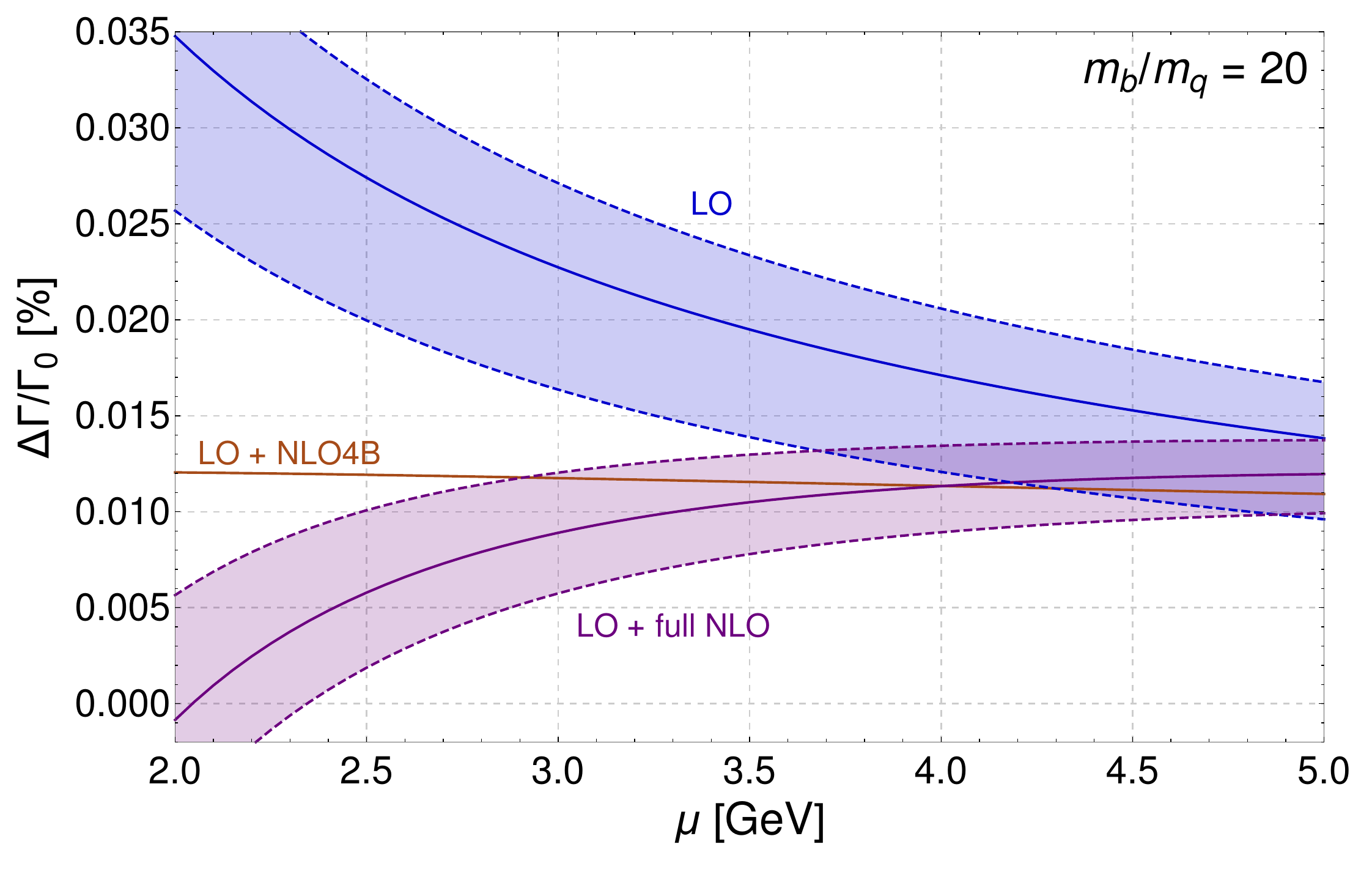}
    \caption{Dependence of the normalised decay rate $\Delta\Gamma/\Gamma_0$ on the renormalisation scale $\mu$ at leading order (LO), next-to-leading order including only the four-body contributions calculated in~\citere{Huber:2014nna} (LO~+~NLO4B), and including in addition the   contributions calculated in the present work (LO~+~full~NLO). The uncertainty bands shown for the blue and purple curves include the dependence on $\mu_0$, $m_b$, $z_c$, $m_t$, and $\alpha_s(M_Z)$ added in quadrature. The dependence on $\log(m_b/m_q)$ is reflected by the two panels for $m_b/m_q=50$ (left) and $m_b/m_q=20$ (right). See text for further details. \label{fig:plotsmu}}    
\end{figure}
%%%%%%%%%%%%%%%%%%%%%%%%%%%%%%%%
One observes that for $m_b/m_q=50$ the NLO4B+5B term is as large as the NLO4B one, while it is significantly smaller for $m_b/m_q=20$. This can also be seen from the plots in figure~\ref{fig:plotsmu} where we show the dependence of $\Delta\Gamma/\Gamma_0$ on the renormalisation scale $\mu$ for $\delta = 1-2E_0/m_b=0.3305$ and two different values of the collinear logarithm $\log(m_b/m_q) =\log(50)$ (left panel) and $\log(m_b/m_q) =\log(20)$ (right panel). The uncertainty bands shown for the blue (LO) and purple (LO~$+$~full NLO) curves were obtained by varying $\mu_0$, $m_b$, $z_c$, $m_t$, and $\alpha_s(M_Z)$ in the ranges indicated in table~\ref{tab:inputs}, and adding the individual pieces in quadrature, bearing in mind that the LO does neither depend on $z_c$ nor on $m_t$. The size of the uncertainty bands is in general smaller for NLO compared to LO, as expected. The bulk of the uncertainty stems from the variation of $\mu_0$ at LO and $z_c$ at NLO (we comment on the latter in next paragraph). The aforementioned RG invariance is reflected by the fact that the purple curves are in general flatter than their blue counterparts, in particular in the region $\mu>3.5$~GeV. One observes in addition that for certain values of $\mu$ and $\log(m_b/m_q)$ the brown curves (LO~$+$~NLO4B) are even flatter than the purple ones. However, we consider this behaviour to be accidental; only the LO~$+$~full NLO result is formally RG invariant to the order we are working.

%%%%%%%%%%%%%%%%%%%%%%%%%%%%%%%%
\begin{figure}[t]
    \centering
    \includegraphics[width=0.483\textwidth]{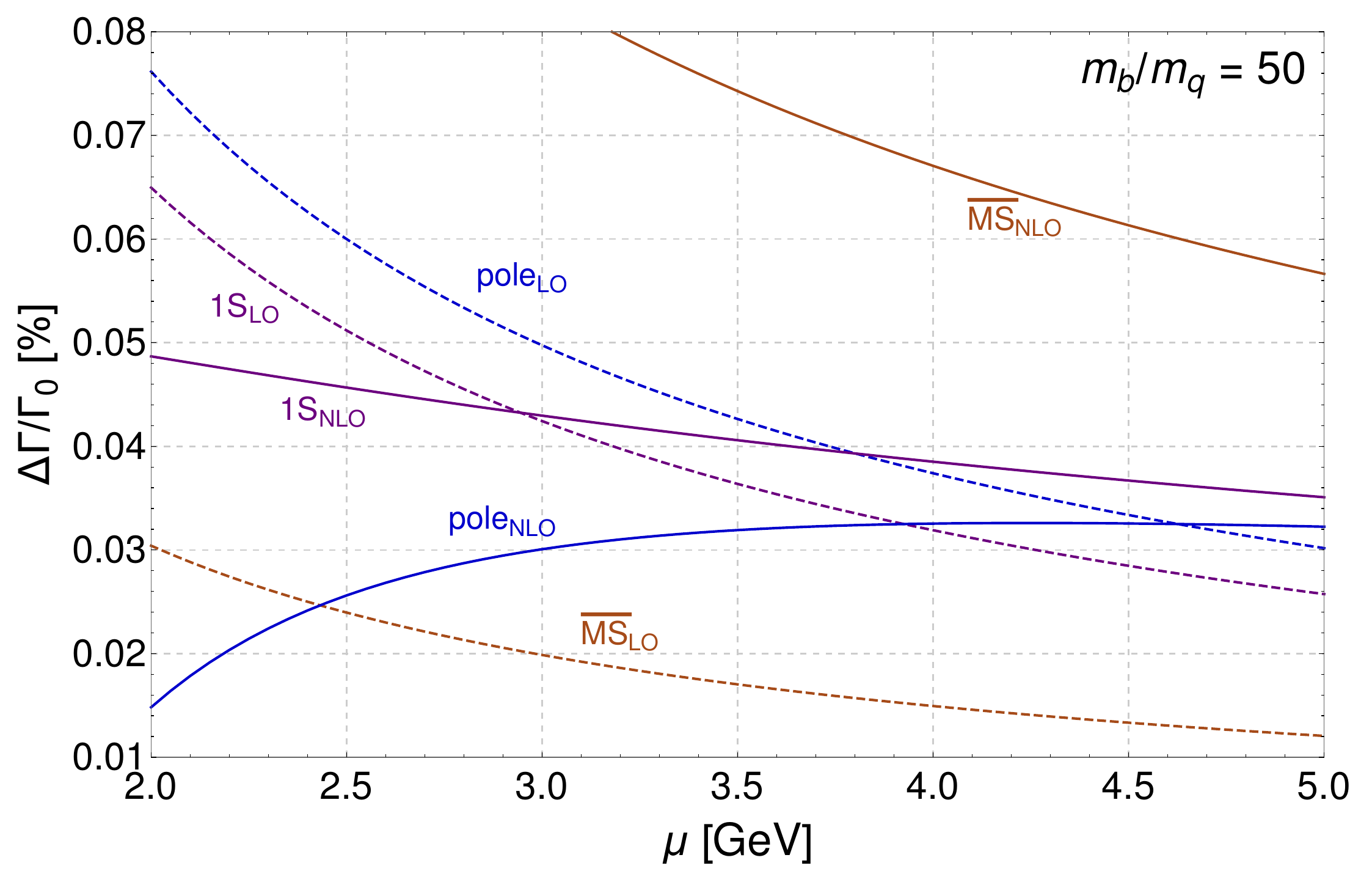}
    \includegraphics[width=0.49\textwidth]{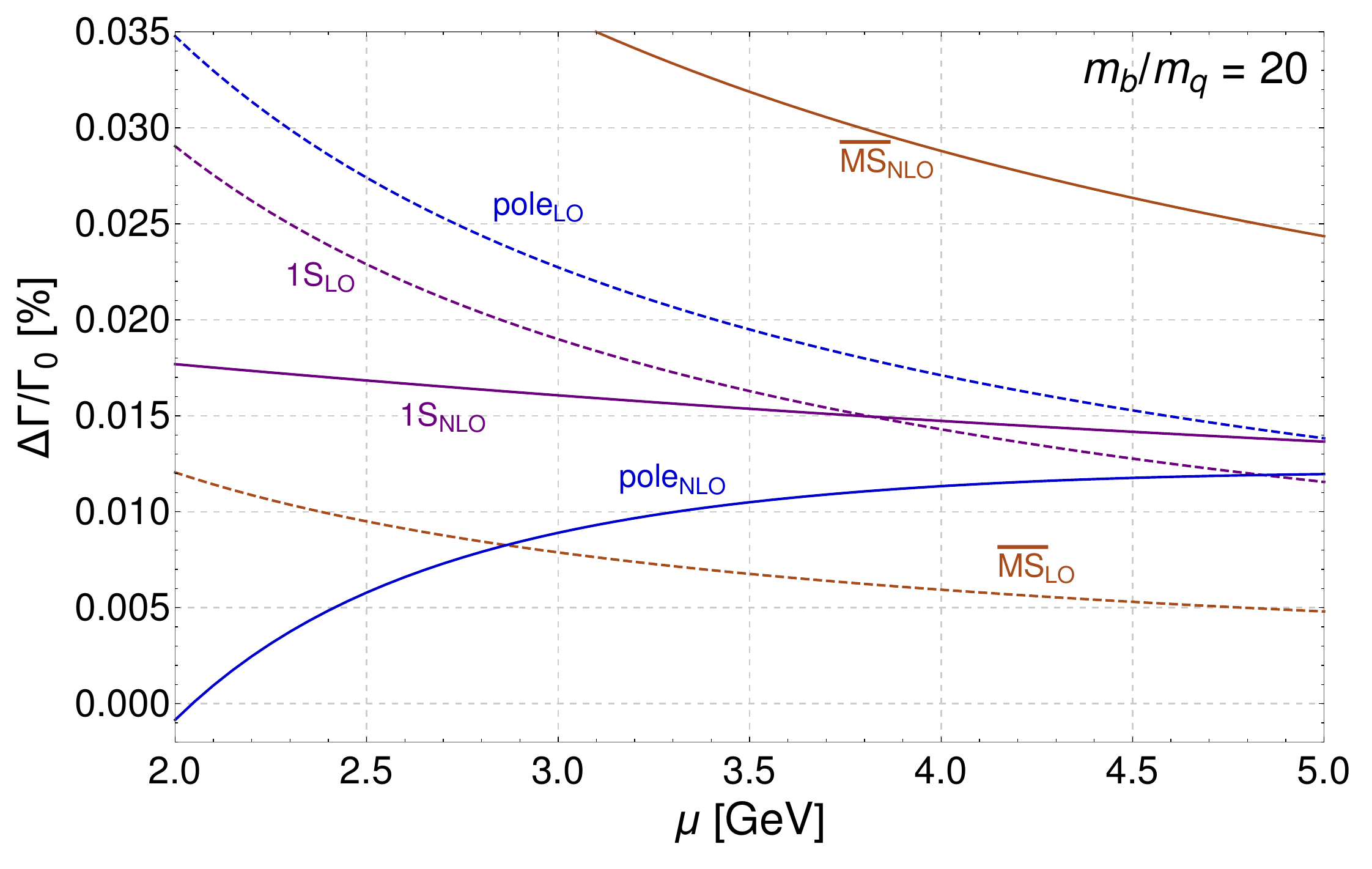}
    \caption{Scheme dependence of the normalised decay rate $\Delta\Gamma/\Gamma_0$ as a function of the renormalisation scale $\mu$ at LO (dashed) and NLO (solid) in the pole (blue), 1S (purple) and $\overline{\text{MS}}$ (brown) scheme. See text for further details. \label{fig:plotsschememu}}    
\end{figure}
%%%%%%%%%%%%%%%%%%%%%%%%%%%%%%%%

Our default renormalisation scheme for the bottom and charm mass is the pole scheme. The latter appears only from NLO onwards, and hence an existing scheme dependence will only be tamed starting from $\ord{\alpha_s^2}$. To account for this circumstance we vary $z_c$ in the range $z_c=0.12\pm 0.03$ (rather than $\pm 0.01$ as na\"ive error propagation would suggest) to capture the different values of $z_c$ within different schemes. The bottom-quark mass appears already at LO and can be converted perturbatively to the 1S or the $\overline{\text{MS}}$ scheme using~\cite{Tarrach:1980up,Pineda:1997hz,hoang:2000fm}
\begin{eqnarray}
 m_b^{\text{pole}} &=& \overline{m}_b(\overline{m}_b) \, \left(1 + \frac{16}{3} \frac{\alpha_s(\overline{m}_b)}{4\pi} + \ldots\right) \,  , \label{eq:poleMSbar}\\[0.3em]
 m_b^{\text{pole}} &=& m_b^{\text{1S}}  \, \left(1 + \frac{C_F^2}{8} \, \alpha_s^2(\mu) + \ldots \right) \, .  \label{eq:pole1S}
\end{eqnarray}
Note that both correction terms in the parentheses are treated as first-order corrections~\cite{hoang:2000fm}. We convert the pole mass of the bottom quark perturbatively to the 1S and $\overline{\text{MS}}$ mass, respectively, and study the renormalisation scheme dependence of the heavy quarks $b$ and $c$ in figure~\ref{fig:plotsschememu} to lowest and first order in $\alpha_s$. We observe that the pole and the 1S scheme yield in general similar results and the scheme dependence is reduced at NLO. The $\overline{\text{MS}}$ scheme, on the other hand, is quite different from the other two, both, at LO and NLO. A possible explanation is two-fold. First, the dependence on the charm-quark mass enters only starting from NLO, and hence a scheme dependence of the observed size is not entirely surprising (note that for the charm quark there is no 1S mass and hence we use the default value $z_c=0.12$ in that case). Second, in eqs.~\eqref{eq:poleMSbar} and~\eqref{eq:pole1S} the first-order correction is a better numerical approximation for the 1S compared to the $\overline{\text{MS}}$ scheme. A more detailed analysis of this behaviour is, however, beyond the scope of this paper. To be meaningful, such an investigation needs to be embedded into other, already existing, higher-order corrections to $\bar B \to X_s \gamma$.

In figure~\ref{fig:plotsE0}, we plot the dependence of $\Delta\Gamma/\Gamma_0$ on the photon-energy cut $E_0$, again for $m_b/m_q =50$ (left panel) and $m_b/m_q =20$ (right panel). This time, the uncertainty bands comprise the variation of $\mu$, $\mu_0$, $m_b$, $z_c$, $m_t$, and $\alpha_s(M_Z)$ added in quadrature. Note that here we vary the scales $\mu$ and $\mu_0$ individually, and include only the larger of the two variations as overall scale-uncertainty into the error budget. The uncertainty bands are in general more narrow for NLO compared to LO. Finally, figure~\ref{fig:plotsschemeE0} displays the renormalisation scheme dependence of the heavy quarks as a function of $E_0$. As in the case of the scale-dependence, the largest variation is observed when using the $\overline{\text{MS}}$ mass. We emphasise again that a thorough investigation of this points will require the inclusion of other higher-order corrections to $\bar B \to X_s \gamma$, which is beyond the subject of the present paper.

%%%%%%%%%%%%%%%%%%%%%%%%%%%%%%%%
\begin{figure}[t]
    \centering
    \includegraphics[width=0.483\textwidth]{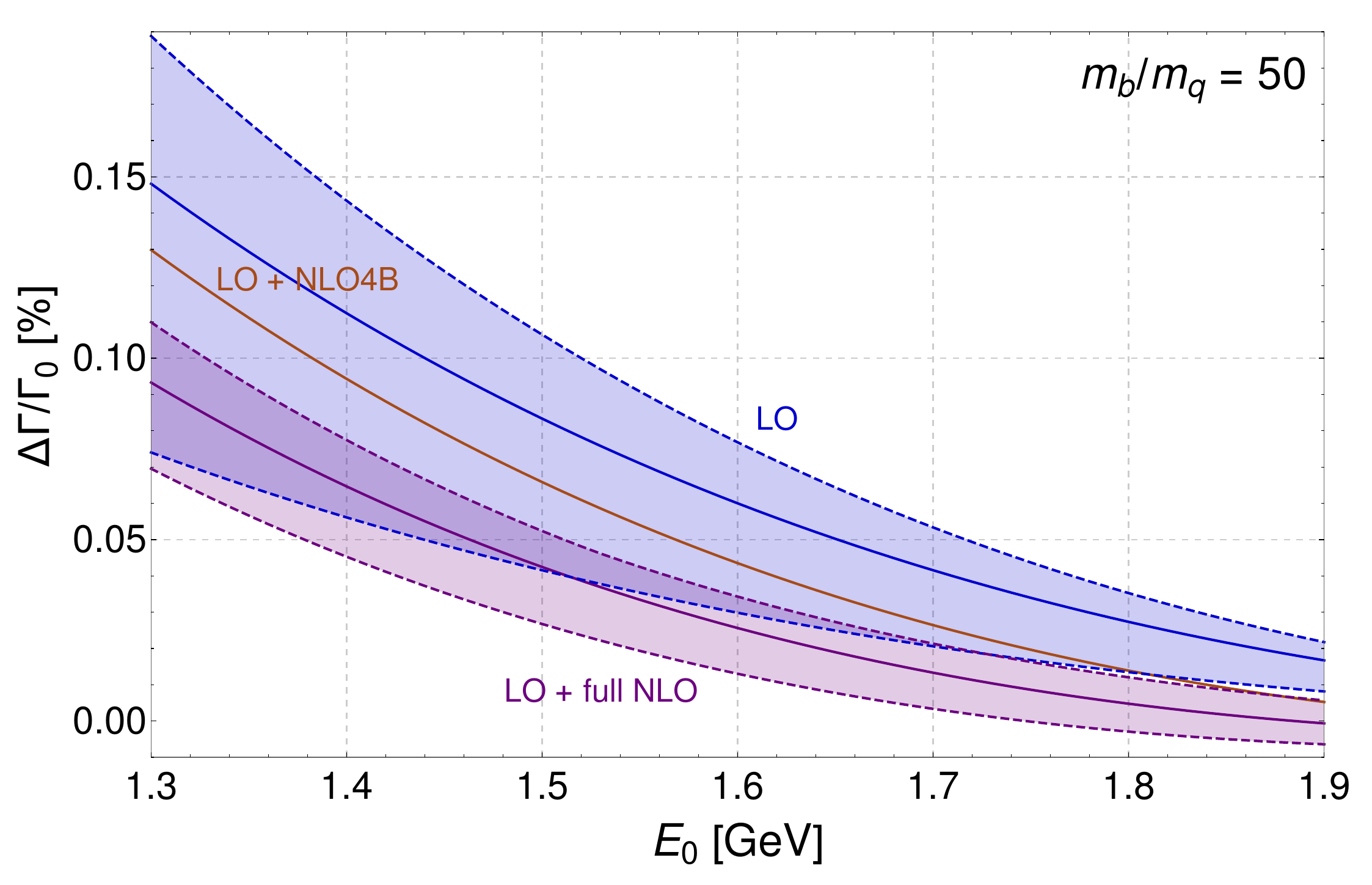}
    \includegraphics[width=0.49\textwidth]{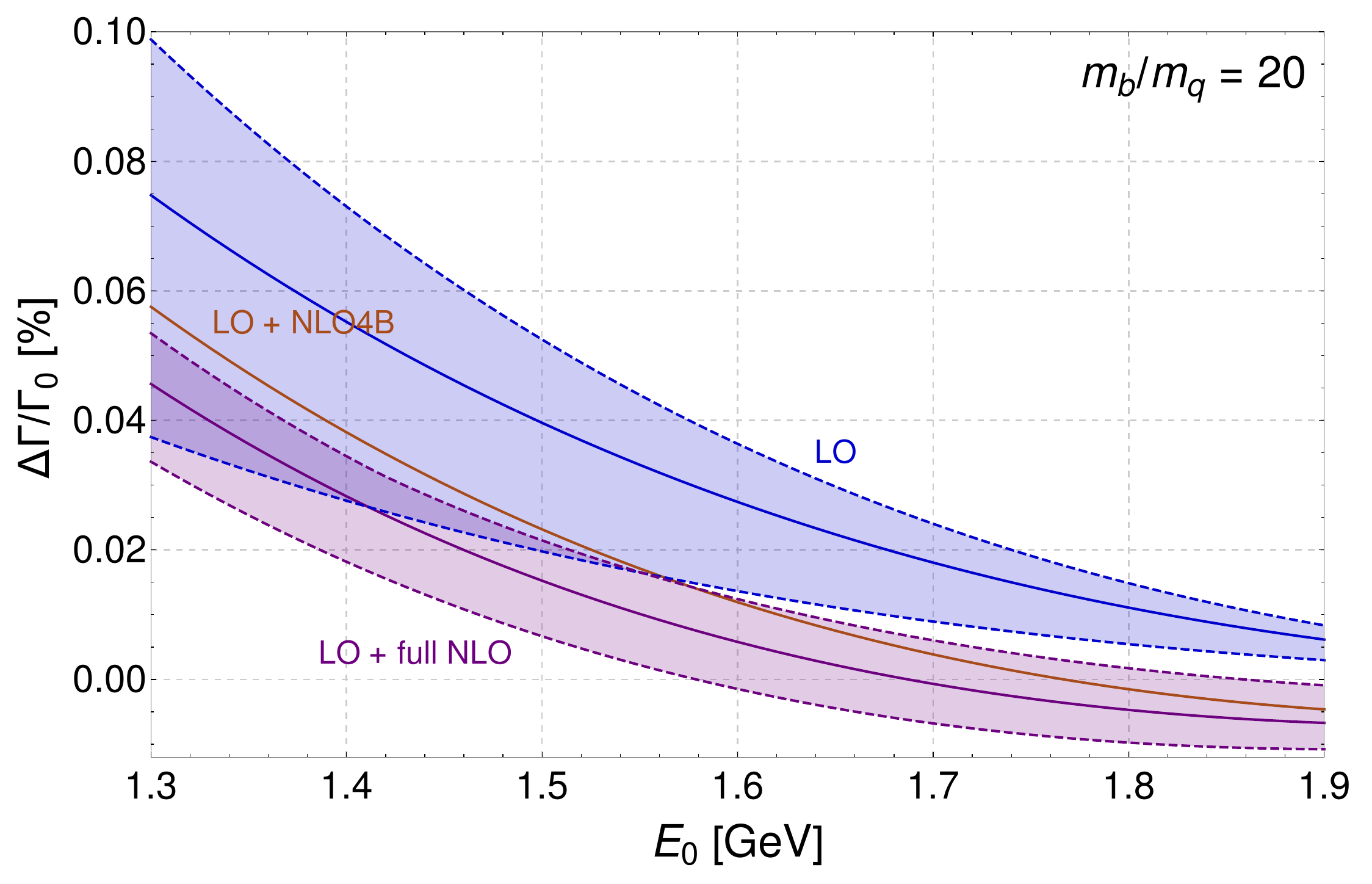}
    \caption{Dependence of the normalised decay rate $\Delta\Gamma/\Gamma_0$ on the photon-energy cut $E_0$ at LO (blue), LO~+~NLO4B (brown), and LO~+~full~NLO (purple). The uncertainty bands shown for the blue and purple curves include the dependence on scales, $m_b$, $z_c$, $m_t$, and $\alpha_s(M_Z)$ added in quadrature. The dependence on $\log(m_b/m_q)$ is reflected by the two panels for $m_b/m_q=50$ (left) and $m_b/m_q=20$ (right). See text for further details.\label{fig:plotsE0}}    
\end{figure}
%%%%%%%%%%%%%%%%%%%%%%%%%%%%%%%%

\section{Conclusion}
\label{sec:conclusion}

The CP- and isospin averaged branching ratio of the inclusive radiative decay $\bar{B}\to X_s \gamma$ represents one of the most suitable observables in quark-flavour physics for precise determinations both, on the experimental and theoretical side. The foreseen improvement in precision on the experimental side --~the anticipated uncertainty at the end of Belle II is $\pm2.6\%$~\cite{Belle-II:2018jsg,Ishikawa:2019TalkLyon}~-- justifies every effort to reduce the uncertainty also on the theoretical side.

In the present article we calculate the last missing contributions to formally complete $\bar{B}\rightarrow X_s \gamma$ at NLO at the leading power. These amount to computing those interferences between current-current and QCD penguin operators in which the one-loop four-body contributions $b\rightarrow s q \bar{q} \gamma$ to  $\Gamma(\bar{B}\rightarrow X_s \gamma)$ have to be supplemented by the corresponding tree-level five-body bremsstrahlung $b\rightarrow s q \bar{q} \gamma g$ contributions. The smallness of the Wilson coefficients of penguin operators and CKM-suppression of current-current operators suggests that this contribution should be small. However, only an explicit calculation can validate this estimate into a proper statement. The calculation is technically involved in several respects. First, in products of traces involving $\gamma_5$ we apply the so-called KKS scheme, which is a reading point scheme that allows anti-commutativity but forbids to use cyclicity of the trace. Second, we perform an integral reduction which results in about sixty master integrals that we compute analytically by explicit phase-space integration in dimensional regularisation and the method of differential equations. Moreover, the cancellation of poles in the dimensional regulator $\eps$ is only achieved after proper UV and IR renormalisation. The procedure of renormalising the IR divergences gives rise to collinear logarithms $\log(m_b^2/m_q^2)$ when turning the dimensional into a mass regulator. The origin of the logarithms are from phase-space regions where an energetic collinear photon radiates from a light massive quark in the final state. Finally, all results are obtained in analytic form in terms of multiple polylogarithms. For more details on the computational steps we refer to~\cite{Moos:2024ogv}.

%%%%%%%%%%%%%%%%%%%%%%%%%%%%%%%%
\begin{figure}[t]
    \centering
    \includegraphics[width=0.483\textwidth]{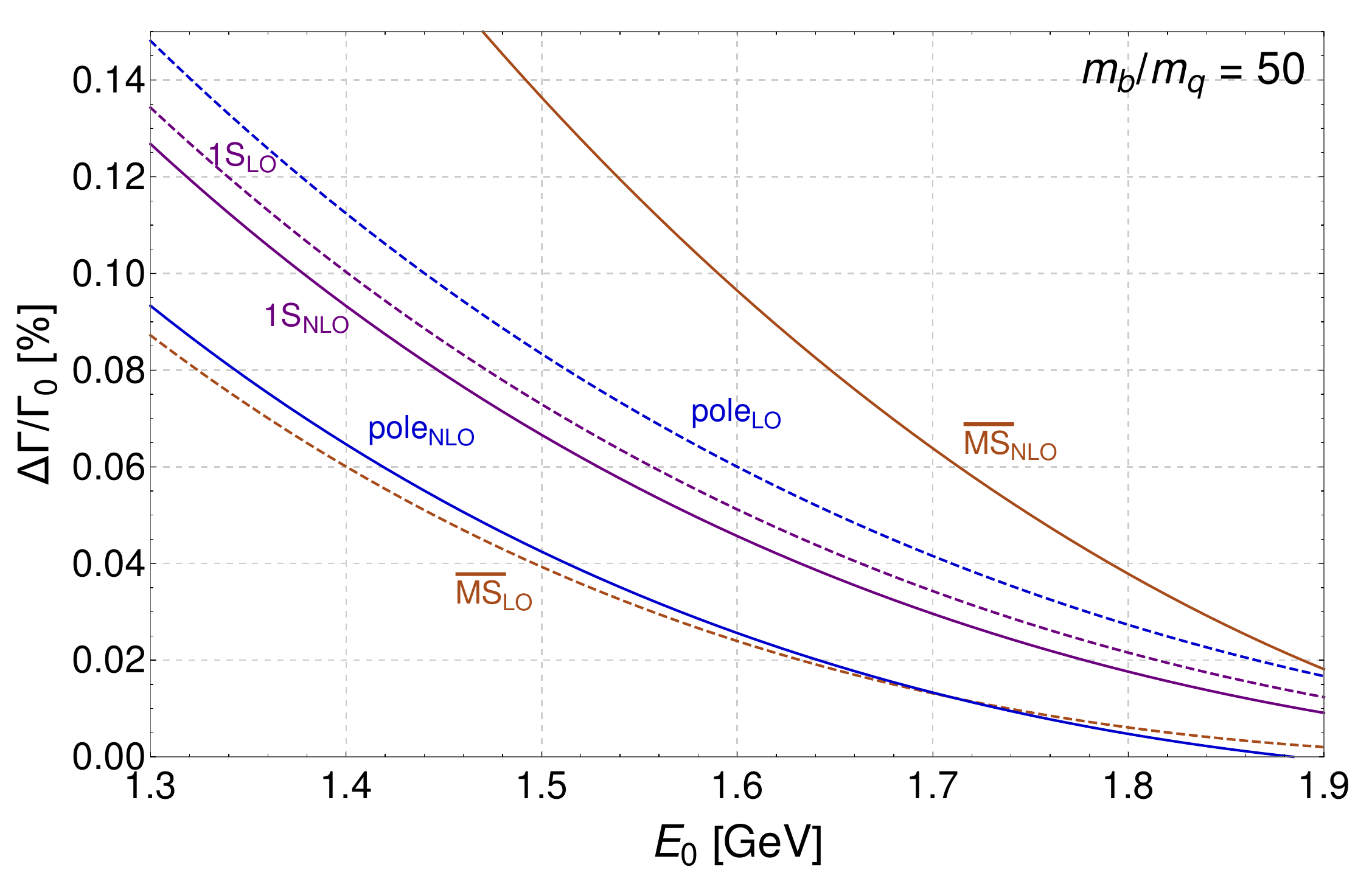}
    \includegraphics[width=0.49\textwidth]{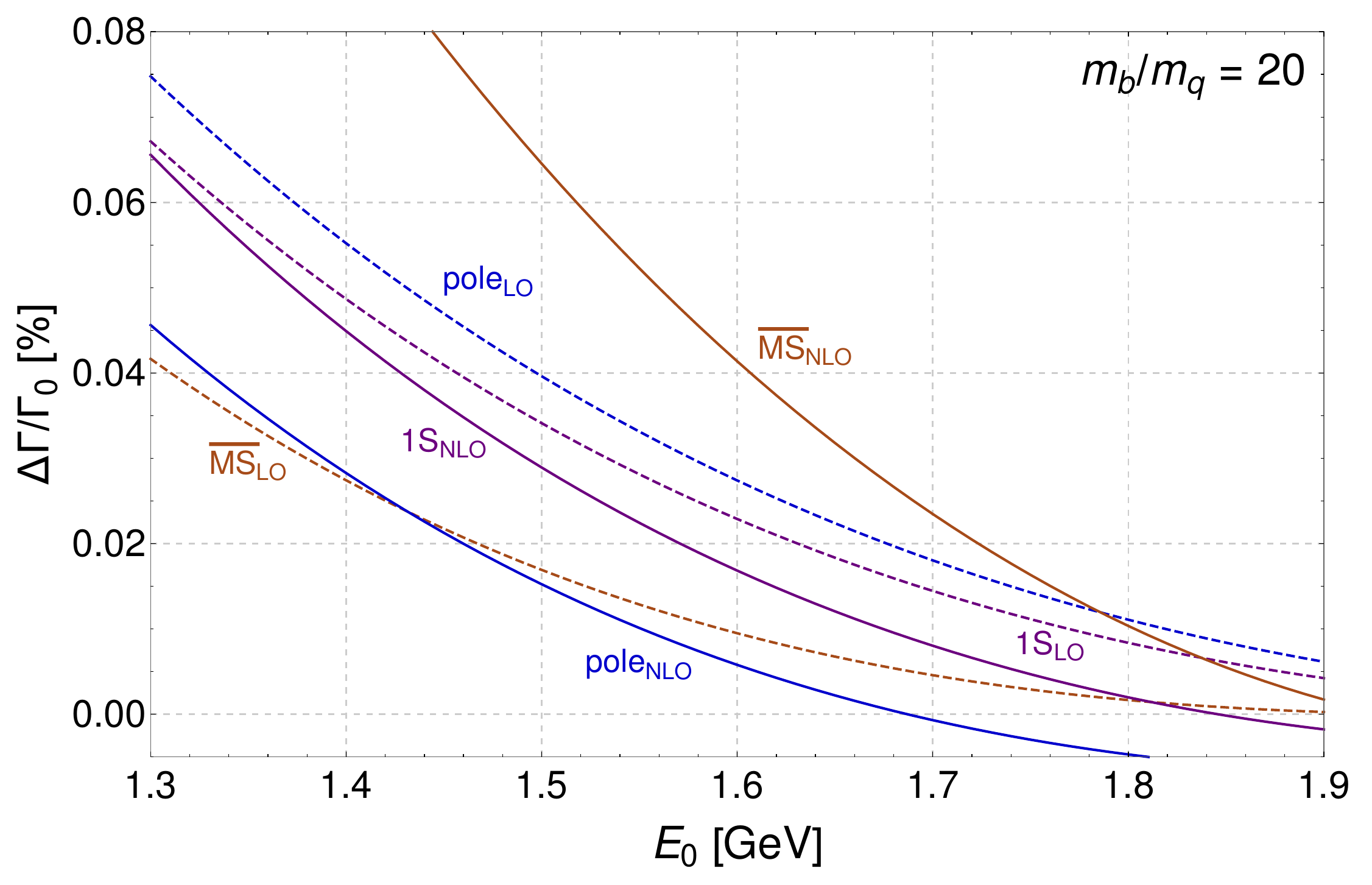}
    \caption{Scheme dependence of the normalised decay rate $\Delta\Gamma/\Gamma_0$ as a function of the photon-energy cut $E_0$ at LO (dashed) and NLO (solid) in the pole (blue), 1S (purple) and $\overline{\text{MS}}$ (brown) scheme. In the two panels $m_b/m_q=50$ (left) and $m_b/m_q=20$ (right) has been chosen. \label{fig:plotsschemeE0}}    
\end{figure}
%%%%%%%%%%%%%%%%%%%%%%%%%%%%%%%%

We find that the numerical effects from the missing multi-parton contributions compared to ones presented in~\citere{Huber:2014nna} are of a similar size, at least for values of $m_b/m_q=50$. Both, the NLO contributions from~\citere{Huber:2014nna} and the present ones reduce the LO contribution and hence lead to a partial cancellation between LO and NLO terms. This partial cancellation renders the overall numerical impact of the multi-body contributions to the $\bar{B}\to X_s \gamma$ decay rate small, the precise numerical values are given in section~\ref{sec:numericalresults}.

%%%%%%%%%%%%%%%%%%%%%%%%%%%%%%%%%%%%%%%%%%%%%%%%%%%%%%%%

\section*{Acknowledgements}

During the course of this work we benefitted from discussions and useful insights from many colleagues. We are indebted to Guido Bell, Robin Br\"user, Thomas Gehrmann, Vitaly Magerya, Vicent Mateu, Kirill Melnikov, Miko{\l}aj Misiak, German Sborlini, and Maximilian Stahlhofen for insightful discussions and explanations. Moreover, we would like to thank Matteo Fael, Martin Gorbahn, Uli Haisch, Florian Herren, Stephen Jones, Vlad Shtabovenko, Javier Virto and Zac W\"uthrich for useful discussions and correspondence. The calculations of this work were partially performed on the HPC cluster {\texttt{OMNI}} at Siegen University. This research was supported by the Deutsche Forschungsgemeinschaft (DFG, German Research Foundation) under grant 396021762 --- TRR 257 ``Particle Physics Phenomenology after the Higgs Discovery''. KB thanks the university of Salamanca for hospitality while parts of this work were completed. The Feynman diagrams were drawn with the help of Axodraw~\cite{Vermaseren:1994je} and JaxoDraw~\cite{Binosi:2003yf}.

%%%%%%%%%%%%%%%%%%%%%%%%%%%%%%%%%%%%%%%%%%%%%%%%%%%%%%%%

\begin{appendix}

%%%%%%%%%%%%%%%%%%%%%%%%%%%%%%%%%%%%%%%%%%%%%%%%%%%%%%%%

\section{Integral relations}
\label{app:integralrelations}

In this appendix we collect relations between phase-space integrals. In the case of the four-body integrals and with $m_b$ set to unity, they read
\begin{align}
F_{4\mathrm{B}13} &= \frac{(1-\epsilon )}{z-1}F_{4\mathrm{B}8}+\frac{(2 z (\epsilon -1)+2 \epsilon -1)}{z-1}F_{4\mathrm{B}12}+\frac{2 z (4 \epsilon -3)}{(z-1)^2}F_{4\mathrm{B}24} \,,\\[0.4em]
F_{4\mathrm{B}16} &= \frac{(\epsilon -1) }{\bar{z}^2 \epsilon }F_{4\mathrm{B}8}+\frac{(2-4 \epsilon ) }{\bar{z} \epsilon }F_{4\mathrm{B}9}+\frac{\epsilon}{2 \epsilon-1}F_{4\mathrm{B}14}-\frac{1}{\bar{z} \epsilon }F_{4\mathrm{B}11} \nonumber \\[0.4em]
   &+\frac{(6 \bar{z} \epsilon -4 \bar{z}-4 \epsilon +3)}{\bar{z}^2 \epsilon }F_{4\mathrm{B}12}-\frac{2 (\bar{z}-1) (4 \epsilon -3)}{\bar{z}^3 \epsilon }F_{4\mathrm{B}24}+\frac{\epsilon}{1-2 \epsilon }F_{4\mathrm{B}15} \,,  \\
F_{4\mathrm{B}26} &=\frac{(\epsilon -1) (z (4 \epsilon -3)+1)}{(z-1)^3 \epsilon }F_{4\mathrm{B}8}+\frac{(2-4 \epsilon )}{z-1}F_{4\mathrm{B}9}+\frac{\left(\epsilon -3 \epsilon ^2\right)}{2 \epsilon
   -1}F_{4\mathrm{B}14} \nonumber \\[0.4em]
   &+\frac{1}{z-1}F_{4\mathrm{B}10}+\frac{(1-2 \epsilon )}{\epsilon -z \epsilon }F_{4\mathrm{B}11}+\frac{\epsilon  (3 \epsilon -1)}{2 \epsilon -1} F_{4\mathrm{B}15} \nonumber \\
   &+\frac{\left(2 \left(z^2-12 z+3\right) \epsilon ^2+\left(2 z^2+23 z-5\right) \epsilon -2 z^2-5 z+1\right)}{(z-1)^3 \epsilon }F_{4\mathrm{B}12}\\
   &-\frac{2 z (4 \epsilon -3) ((z+3) \epsilon -z-1) }{(z-1)^4 \epsilon }F_{4\mathrm{B}24} \nonumber \,.
\end{align}
Finally, in the case of the five-body integrals, they read (we set again $m_b=1$)
\begin{align}
F_{\text{5B8}} &=
\left[
\frac{7 (4 {\bar z}-3)}{30 (5 \eps-2) ({\bar z}-1) ({\bar z}+1)}
+\frac{(6 {\bar z}+7)}{6 \eps ({\bar z}-1) ({\bar z}+1)}
+\frac{\left(5 {\bar z}^2-2 {\bar z}-4\right)}{6 (2 \eps-1) ({\bar z}-1) ({\bar z}+1)} \right. \nonumber\\[0.3em]
& \left. \qquad -\frac{\left(5 {\bar z}^2+6 {\bar z}-2\right)}{5 ({\bar z}-1) ({\bar z}+1)}
-\frac{1}{2 \eps^2 ({\bar z}-1) ({\bar z}+1)}
-\frac{1}{6 (2 \eps-1)^2}\right] \, F_{\text{5B1}} +\left[
-\frac{{\bar z}}{6 (2\eps-1)} \right. \nonumber\\[0.3em]
& \qquad \left.
-\frac{\left(3 {\bar z}^2-2 {\bar z}+3\right) }{60 (5 \eps-2) ({\bar z}-1) ({\bar z}+1)}
+\frac{\left(2 {\bar z}^3+{\bar z}^2+4 {\bar z}-3\right) }{12 \eps ({\bar z}-1) ({\bar z}+1)}
-\frac{\left(10 {\bar z}^3+11 {\bar z}^2+11 {\bar z}-14\right)}{15 ({\bar z}-1) ({\bar z}+1)}
\right] \, F_{\text{5B2}} \nonumber\\[0.3em]
& +\left[
\frac{({\bar z}-1) {\bar z}}{12 (2 \eps-1) ({\bar z}+1)}
+\frac{({\bar z}-1) {\bar z}}{3 (5 \eps-2) ({\bar z}+1)}
+\frac{({\bar z}-1) {\bar z}}{8 (2 \eps-1)^2 ({\bar z}+1)}
+\frac{({\bar z}-1) {\bar z}}{8 ({\bar z}+1)}\right] \, F_{\text{5B3}} \nonumber\\[0.3em]
& +\left[
\frac{({\bar z}-1) {\bar z}}{(2 \eps-1) ({\bar z}+1)}
-\frac{5 ({\bar z}-1) {\bar z}}{24 (4 \eps-1) ({\bar z}+1)}
-\frac{7 ({\bar z}-1) {\bar z}}{3 (5 \eps-2) ({\bar z}+1)}
-\frac{3 ({\bar z}-1) {\bar z}}{8 (2 \eps-1)^2 ({\bar z}+1)}\right] \, F_{\text{5B4}} \nonumber\\[0.3em]
& +\left[
-\frac{(3 {\bar z}-1) (5 {\bar z}-9)}{60 (5 \eps-2) ({\bar z}-1)}
+\frac{(2 {\bar z}+1) \left(5 {\bar z}^2+10 {\bar z}-8\right) }{40 ({\bar z}-1)}
+\frac{\left({\bar z}^3-{\bar z}^2+5 {\bar z}-2\right) }{24 (2 \eps-1)^2 ({\bar z}-1)} \right. \nonumber\\[0.3em]
& \qquad \left. +\frac{\left(5 {\bar z}^3+8 {\bar z}^2+{\bar z}-4\right)}{24 (2 \eps-1) ({\bar z}-1)}\right] \,  F_{\text{5B6}}
+\left[
-\frac{({\bar z}+1) {\bar z}}{4 (2 \eps-1)}
-\frac{({\bar z}+1)  {\bar z}}{3 (5 \eps-2) ({\bar z}-1)}
+\frac{({\bar z}+1) (3 {\bar z}-2)  {\bar z}}{24 (4 \eps-1) ({\bar z}-1)} \right. \nonumber\\[0.3em]
& \qquad \left. -\frac{({\bar z}+1)  {\bar z}^2}{8 (2 \eps-1)^2 ({\bar z}-1)} \right] \, F_{\text{5B7}}
+\left[
-\frac{ {\bar z}}{12 (2 \eps-1)}
+\frac{({\bar z}+1) (2 {\bar z}-1) }{24 \eps ({\bar z}-1)}
-\frac{(5 {\bar z}-3) }{120 (5 \eps-2) ({\bar z}-1)} \right. \nonumber\\[0.3em]
& \qquad \left. -\frac{\left(5 {\bar z}^2-2\right) }{10 ({\bar z}-1)}\right] \,  F_{\text{5B9}}
+\left[
+\frac{({\bar z}-1)  {\bar z}}{8 (2 \eps-1)}
-\frac{({\bar z}-1) {\bar z}}{48 (4 \eps-1)}
-\frac{8 ({\bar z}-1)  {\bar z}}{15 (5 \eps-2)}
-\frac{({\bar z}-1)  {\bar z}}{8 (2 \eps-1)^2} \right. \nonumber\\[0.3em]
& \qquad \left. -\frac{3}{80} ({\bar z}-1)  {\bar z} \right] \, F_{\text{5B11}} \, , \\[0.6em]
F_{\text{5B33}} & =
\left[
-\frac{25 (107 {\bar z}+10) \left(3 {\bar z}^2-201 {\bar z}+89\right)}{749112 (7 \eps-4){\bar z}}
+\frac{\left(4 {\bar z}^2-68 {\bar z}+19\right) }{160 (2 \eps-1)}
+\frac{\left({\bar z}^3-7 {\bar z}^2+5 {\bar z}-2\right) }{30 \eps^2{\bar z}}\right. \nonumber\\[0.3em]
& \qquad \left.
-\frac{3 ({\bar z}+26)\left(278 {\bar z}^2-1889 {\bar z}-36\right) }{660275 (7 \eps-5){\bar z}}
+\frac{11 \left(25 {\bar z}^3-297 {\bar z}^2+1798 {\bar z}-844\right) }{36015 (7 \eps-3){\bar z}}\right. \nonumber\\[0.3em]
& \qquad \left.
-\frac{\left(251 {\bar z}^3-1587 {\bar z}^2+2475 {\bar z}-1162\right) }{1800 \eps {\bar z}}
-\frac{8 \left(1443 {\bar z}^3-181 {\bar z}^2-1332 {\bar z}-1100\right) }{324135 (7\eps-6) {\bar z}}\right. \nonumber\\[0.3em]
& \qquad \left.
+\frac{\left(5400 {\bar z}^3-32694 {\bar z}^2+75195 {\bar z}-53504\right)}{115248 {\bar z}}\right. \nonumber\\[0.3em]
& \qquad \left.
+\frac{\left(6900 {\bar z}^3-11392 {\bar z}^2-61767 {\bar z}+111616\right)}{617760 (2 \eps-3) {\bar z}} \right] \, F_{\text{5B1}}
+\left[
-\frac{5 \left(4 {\bar z}^2+12 {\bar z}-5\right) }{288 (4\eps-3)}\right. \nonumber\\[0.3em]
& \qquad \left.
-\frac{({\bar z}+1) \left(9 {\bar z}^2+4 {\bar z}-2\right) }{225 (3 \eps-2){\bar z}}
-\frac{5 \left(3 {\bar z}^2-201 {\bar z}+89\right) \left(15 {\bar z}^2+31 {\bar z}-10\right)}{374556 (7 \eps-4) {\bar z}}\right. \nonumber\\[0.3em]
& \qquad \left.
-\frac{({\bar z}-1) \left({\bar z}^3-7 {\bar z}^2+5{\bar z}-2\right) }{180 \eps {\bar z}}
+\frac{27 ({\bar z}+26) \left(16 {\bar z}^3+156{\bar z}^2-39 {\bar z}+12\right) }{132055 (7 \eps-5) {\bar z}}\right. \nonumber\\[0.3em]
& \qquad \left.
-\frac{11 (4 {\bar z}-1) \left(25{\bar z}^3-297 {\bar z}^2+1798 {\bar z}-844\right) }{540225 (7 \eps-3) {\bar z}}\right. \nonumber\\[0.3em]
& \qquad \left.
-\frac{20\left(33 {\bar z}^4-495 {\bar z}^3+1100 {\bar z}^2-790 {\bar z}+220\right) }{64827 (7 \eps-6){\bar z}}\right. \nonumber\\[0.3em]
& \qquad \left.
+\frac{\left(6480 {\bar z}^4-41004 {\bar z}^3+54944 {\bar z}^2-58715 {\bar z}+53504\right)}{691488 {\bar z}}\right. \nonumber\\[0.3em]
& \qquad \left.
-\frac{\left(8400 {\bar z}^4-35112 {\bar z}^3+48746 {\bar z}^2-60871{\bar z}+111616\right) }{1544400 (2 \eps-3) {\bar z}}\right] \, F_{\text{5B2}}
+\left[
-\frac{5 ({\bar z}-1) }{18 (4\eps-3)}\right. \nonumber\\[0.3em]
& \qquad \left.
-\frac{5 ({\bar z}-16) \left(3 {\bar z}^2-201 {\bar z}+89\right) }{93639 (7\eps-4)}
-\frac{18 ({\bar z}+26) \left(9 {\bar z}^2-162 {\bar z}+197\right) }{132055 (7\eps-5)}\right. \nonumber\\[0.3em]
& \qquad \left.
-\frac{\left({\bar z}^3-7 {\bar z}^2+5 {\bar z}-2\right) }{45 \eps}
+\frac{11 \left(25{\bar z}^3-297 {\bar z}^2+1798 {\bar z}-844\right) }{108045 (7 \eps-3)}\right. \nonumber\\[0.3em]
& \qquad \left.
-\frac{4 \left(150{\bar z}^3-802 {\bar z}^2+1463 {\bar z}-844\right) }{19305 (2 \eps-3)}
+\frac{16 \left(150{\bar z}^3-725 {\bar z}^2-650 {\bar z}+697\right) }{64827 (7 \eps-6)}\right. \nonumber\\[0.3em]
& \qquad \left.
+\frac{\left(540{\bar z}^3-3156 {\bar z}^2+5395 {\bar z}-2829\right) }{14406}\right] \, F_{\text{5B13}}
+\left[
\frac{50 (4 {\bar z}-1) \left(3{\bar z}^2-201 {\bar z}+89\right) }{93639 (7 \eps-4)}\right. \nonumber\\[0.3em]
& \qquad \left.
-\frac{8 \left(9 {\bar z}^2+4{\bar z}-2\right) }{225 (3 \eps-2)}
+\frac{36 ({\bar z}+26) \left(15 {\bar z}^2+10{\bar z}-3\right) }{26411 (7 \eps-5)}
-\frac{352 \left(5 {\bar z}^3-30 {\bar z}^2+18{\bar z}-5\right) }{21609 (7 \eps-6)}\right. \nonumber\\[0.3em]
& \qquad \left.
+\frac{11 \left(25 {\bar z}^3-297 {\bar z}^2+1798{\bar z}-844\right) }{180075 (7 \eps-3)}
-\frac{\left(81 {\bar z}^3-1368 {\bar z}^2+2827{\bar z}-1565\right)  }{21609}\right. \nonumber\\[0.3em]
& \qquad \left.
+\frac{2 \left(175 {\bar z}^3-969 {\bar z}^2+2001 {\bar z}-1273\right)}{32175 (2 \eps-3)}\right] \, F_{\text{5B15}}
+\left[
\frac{216 ({\bar z}-1) ({\bar z}+26) \left(2 {\bar z}^2-{\bar z}+1\right)}{26411 (7 \eps-5)}\right. \nonumber\\[0.3em]
& \qquad \left.
-\frac{50 ({\bar z}-2) ({\bar z}-1) \left(3 {\bar z}^2-201 {\bar z}+89\right)}{93639 (7 \eps-4)}
-\frac{880 ({\bar z}-2) ({\bar z}-1) \left(3 {\bar z}^2+2 {\bar z}-2\right)  }{21609 (7 \eps-6)}\right. \nonumber\\[0.3em]
& \qquad \left.
-\frac{4 ({\bar z}-1) \left(9 {\bar z}^2+4 {\bar z}-2\right)}{225 (3 \eps-2)}
-\frac{({\bar z}-1) (9 {\bar z}-11) \left(45 {\bar z}^2-187 {\bar z}+152\right)}{21609}\right. \nonumber\\[0.3em]
& \qquad \left.
-\frac{22 ({\bar z}-1) \left(25 {\bar z}^3-297 {\bar z}^2+1798 {\bar z}-844\right)}{180075 (7 \eps-3)}\right. \nonumber\\[0.3em]
& \qquad \left.
+\frac{2 ({\bar z}-1) \left(525 {\bar z}^3-2282 {\bar z}^2+3453{\bar z}-1744\right) }{32175 (2 \eps-3)}\right] \, F_{\text{5B16}} \, ,  \\[0.6em]
F_{\text{5B34}} & = 
\left[
\frac{\left(2 {\bar z}^2-2 {\bar z}-1\right) }{8 (2\eps-1) ({\bar z}-1)^2}
-\frac{(2 {\bar z}-5) {\bar z} }{3 (3 \eps-1)({\bar z}-1)^2}
-\frac{({\bar z}-3) {\bar z} }{4 \eps ({\bar z}-1)^2}
+\frac{\left(2{\bar z}^2-32 {\bar z}-3\right) }{24 ({\bar z}-1)^2} \right] \, F_{\text{5B1}} \nonumber\\[0.3em]
&+\left[
\frac{({\bar z}-3) {\bar z}^2 }{12 \eps ({\bar z}-1)^2}
+\frac{\left(28 {\bar z}^3-33 {\bar z}^2+30 {\bar z}+15\right) }{240 (4\eps-3) ({\bar z}-1)^2}
+\frac{(2 {\bar z}-5){\bar z} }{45 (3 \eps-1) ({\bar z}-1)}
-\frac{({\bar z}-4) {\bar z} }{36 (3\eps-2) ({\bar z}-1)} \right. \nonumber\\[0.3em]
& \qquad \left.
+\frac{\left(8 {\bar z}^3+11 {\bar z}^2+98{\bar z}-45\right)}{144 ({\bar z}-1)^2} \right] \,  F_{\text{5B2}}
+\left[
\frac{5 {\bar z}^2}{48 (4 \eps-3) ({\bar z}-1)}
+\frac{({\bar z}-3) {\bar z} }{12 \eps ({\bar z}-1)}\right. \nonumber\\[0.3em]
& \qquad \left.
-\frac{3 ({\bar z}-4) {\bar z} }{16({\bar z}-1)} \right] \, F_{\text{5B22}}
+\left[
-\frac{\left(5 {\bar z}^2+4 {\bar z}-4\right) {\bar z}^2}{16 (4 \eps-3) ({\bar z}-1)^2}
+\frac{\left(3 {\bar z}^2+4 {\bar z}-2\right) {\bar z}}{12 (3 \eps-2) ({\bar z}-1)}\right. \nonumber\\[0.3em]
& \qquad \left.
+\frac{\left(3 {\bar z}^3-16 {\bar z}^2+12{\bar z}-8\right) {\bar z} }{48 ({\bar z}-1)^2} \right] \, F_{\text{5B25}}
+\left[
\frac{3 {\bar z}^4 }{5 (4 \eps-3) ({\bar z}-1)^2}
+\frac{(2 {\bar z}-5) {\bar z}^3 }{15 (3 \eps-1) ({\bar z}-1)^2}\right. \nonumber\\[0.3em]
& \qquad \left.
-\frac{2 {\bar z}^3 }{3 (3 \eps-2) ({\bar z}-1)} \right] \, F_{\text{5B26}}
+\frac{1}{2} F_{\text{5B29}} \, .
\end{align}

\end{appendix}

\bibliographystyle{jhep}
\bibliography{references.bib}

\end{document}